\documentclass[fleqn,usenatbib]{mnras}
\usepackage{multicol}
\usepackage{newtxtext,newtxmath}

\usepackage[T1]{fontenc}

\usepackage{graphicx}

\DeclareRobustCommand{\VAN}[3]{#2}
\let\VANthebibliography\thebibliography
\def\thebibliography{\DeclareRobustCommand{\VAN}[3]{##3}\VANthebibliography}

\usepackage{graphicx}	
\usepackage{amsmath}	
\usepackage[justification=centering]{caption}

\usepackage{hyperref}
\hypersetup{
    colorlinks=true,
    linkcolor=blue,
    filecolor=blue,      
    urlcolor=blue,
    }

\AtBeginDocument{\mathcode`v=\varv}
\newcommand{\NHI}{\ifmmode N_{{\mathrm{H}} \, \mathrm{I}} \else $N_{{\mathrm{H}} \, \mathrm{I}}$\fi} 
\newcommand{\HI}{\ifmmode \mathrm{\ion{H}{I}} \else \ion{H}{I} \fi}

\def\GHz{\ifmmode $\,GHz$\else \,GHz\fi}
\def\MJysr{\ifmmode \,$MJy\,sr\mo$\else \,MJy\,sr\mo\fi}
\def\microns{\ifmmode \,\mu$m$\else \,$\mu$m\fi}

\def\kms{\ifmmode $\,km\,s$^{-1}\else \,km\,s$^{-1}$\fi}

\usepackage{lineno}

\usepackage{orcidlink}

\title[Considerations with stacking \HI\ spectra]{Considerations with stacking absorption spectra: cold \HI\ gas in cirrus region of the Milky Way}

\author[C. Lynn et. al.]{Callum Lynn\orcidlink{0000-0001-6846-5347}$^{1}$,\thanks{E-mail: callum.lynn@anu.edu.au}
Antoine Marchal\orcidlink{0000-0002-5501-232X}$^{1}$,
N.~M. McClure-Griffiths\orcidlink{0000-0003-2730-957X}$^{1}$,
Marc-Antoine Miville-Deschênes\orcidlink{0000-0002-7351-6062}$^{2}$,
\newauthor
Claire E. Murray\orcidlink{0000-0002-7743-8129}$^{3,4}$,
Hiep Nguyen\orcidlink{0000-0002-2712-4156}$^{1}$,
James Dempsey\orcidlink{0000-0002-4899-4169}$^{1,5}$,
Enrico Di Teodoro\orcidlink{0000-0003-4019-0673}$^{6}$,
Jacco Th. van Loon\orcidlink{0000-0002-1272-3017}$^{7}$,
\newauthor
John M. Dickey\orcidlink{0000-0002-6300-7459}$^{8}$,
Min-Young Lee\orcidlink{0000-0002-9888-0784}$^{9,10}$,
Gilles Joncas\orcidlink{0000-0001-7462-4818}$^{11}$,
Yik Ki Ma\orcidlink{0000-0003-0742-2006}$^{1}$,
Nickolas M. Pingel\orcidlink{0000-0001-9504-7386}$^{12}$,
\newauthor
Snežana Stanimirović\orcidlink{0000-0002-3418-7817}$^{12}$,
Ian Kemp\orcidlink{0000-0002-6637-9987}$^{13}$,
Steven Gibson\orcidlink{0000-0002-1495-760X}$^{14}$,
Helga Dénes\orcidlink{0000-0002-9214-8613}$^{15}$
\\
$^{1}$Research School of Astronomy \& Astrophysics, The Australian National University, Canberra ACT 2611, Australia\\
$^{2}$Laboratoire de Physique de l'École Normale Supérieure, ENS, Université PSL, CNRS, Sorbonne Université, Université de Paris, F-75005 Paris, France\\
$^{3}$Space Telescope Science Institute, 3700 San Martin Drive, Baltimore, MD, 21218, USA \\
$^{4}$Department of Physics \& Astronomy, Johns Hopkins University, 3400 N. Charles Street, Baltimore, MD 21218, USA \\
$^{5}$CSIRO Information Management and Technology, GPO Box 1700 Canberra, ACT 2601, Australia \\
$^{6}$Dipartimento di Fisica e Astronomia, Università degli Studi di Firenze, via G. Sansone 1, 50019 Sesto Fiorentino, Firenze, Italy \\
$^{7}$Lennard-Jones Laboratories, School of Chemical and Physical Sciences, Keele University, Keele, Staffordshire ST5 5BG, UK \\
$^{8}$School of Natural Sciences, University of Tasmania, Private Bag 37, Hobart, TAS 7001, Australia \\
$^{9}$Korea Astronomy and Space Science Institute, 776 Daedeok-daero, Yuseong-gu, Daejeon 34055, Republic of Korea \\
$^{10}$Department of Astronomy and Space Science, University of Science and Technology, 217 Gajeong-ro, Daejeon 34113, Republic of Korea \\
$^{11}$Dépt. de physique, de génie physique et d’optique, Centre de recherche en astrophysique du Québec, 1045 avenue de la médecine, Université Laval, Québec, \\ \, \, Québec, Canada G1V 0A6 \\
$^{12}$Department of Astronomy, University of Wisconsin–Madison, 475 N Charter St, Madison, WI 53703, USA \\
$^{13}$International Centre for Radio Astronomy Research (ICRAR), Curtin University, Bentley, WA 6102, Australia \\
$^{14}$Department of Physics \& Astronomy, Western Kentucky University, 1906 College Heights Boulevard, Bowling Green, KY 42101, USA \\
$^{15}$School of Physical Sciences and Nanotechnology, Yachay Tech University, Hacienda San José S/N, 100119, Urcuquí, Ecuador
}

\date{Accepted 2024 December 16. Received 2024 December 6; in original form 2024 July 16}

\pubyear{2025}

\begin{document}
\label{firstpage}
\pagerange{\pageref{firstpage}--\pageref{lastpage}}
\maketitle

\begin{abstract}
We use the Milky Way neutral hydrogen (\HI) absorption and emission spectra from the Galactic Australian Square Kilometre Array Pathfinder (GASKAP) Phase II Pilot survey along with toy models to investigate the effects of stacking multi-component spectra on measurements of peak optical depth and spin temperature. Shifting spectra by the peak in emission, "primary" components shifted to $0$ \kms\ are correctly averaged. Additional components on individual sightlines are averaged with non-centred velocities, producing a broader and shallower "secondary" component in the resulting stack. Peak optical depths and brightness temperatures of the secondary components from stacks are lower limits of their true average values due to the velocity offset of each component. The spin temperature however is well correlated with the truth since the velocity offset of components affects the emission and absorption spectra equally. Stacking $462$ GASKAP absorption-emission spectral pairs, we detect a component with a spin temperature of $1320 \pm 263$ K, consistent with gas from the unstable neutral medium and higher than any previous GASKAP detection in this region. We also stack 2240 pilot survey spectra containing no Milky Way absorption, revealing a primary narrow and secondary broad component, with spin temperatures belonging to the cold neutral medium (CNM). Spatially binning and stacking the non-detections across the plane-of-sky by their distance from CNM absorption detections, the primary component’s optical depth decreases with distance from known locations of cold gas. The spin temperature however remains stable in both components, over an approximate physical plane-of-sky distance of $\sim100$ pc.

\end{abstract}




\section{Introduction}

Neutral hydrogen (\HI) is the most abundant element in the Universe, making it an important tracer for large-scale galactic morphology. Because \HI\ is the building block of stars, understanding its temperature in the current epoch of the interstellar medium (ISM) is crucial to our understanding of the current state of star formation in the Milky Way.

Through a combination of observations and theory, \HI\ has been best described with a bi-stable model with two stable temperature regimes: the warm neutral medium (WNM), $T_s \sim 4000-8000$ K in the Solar neighbourhood, filling most of the \HI\ volume, and the cold neutral medium (CNM), $T_s \sim 25-250$ K, that exists in dense cold clumps enveloped by the WNM. Between these two temperature regimes lies the thermally unstable neutral medium (UNM), a region thought to be produced by the condensation of the WNM via thermal instability \citep{Field_Thermal+1965,Field_Cosmic-Ray+1969,Wolfire+1995,Wolfire+2003,Hennebelle_Dynamical+1999,Hennebelle_Perrault+2000,Audit_Hennebelle+2005}. 

Observationally, the CNM is the easiest of the three phases to detect in absorption, even through low sensitivity observations, due to its typically high optical depth and column density  \citep{Lazareff+1975,Dickey+1977,Crovisier+1978,Heiles_Troland+2003,Mohan+2004a,Roy+2013,Begum+2010}. However, the UNM and WNM, due to their lower opacity, require high sensitivity observations and instruments to be observed in absorption \citep{Carilli+1998,Dwarakanath+2002,Patra+2018}. Therefore, most of their physical properties have been estimated from \HI\ emission alone, decomposed through methods of Gaussian decomposition  \citep{Haud+2000,Nidever+2008,Lindner+2015,Kalberla+2018,Riener+2019,Marchal+2019,Marchal+2021}.

It is only recently with advances in absorption survey sensitivity and field-of-view that we have the number of sightlines to be able to stack Galactic \HI\ absorption spectra to improve their signal-to-noise. Using absorption sources from the first half of the 21-SPONGE survey, \citet{Murray+2014} stacked the spectra from 19 sightlines with Milky Way \HI\ absorption to improve the optical depth sensitivity to the point of detecting the Milky Way's WNM in absorption ($\sigma_\tau \approx 2.6 \times 10^{-4}$). The spin temperature of this component was approximately $7200^{+1800}_{-1200}$ K. With the completed 21-SPONGE survey catalogue, \cite{Murray+2018} conducted a similar stacking process, detecting a spin temperature of $\sim 10^{4}$ K. While this was one of the first steps in using stacking techniques with \HI\ absorption, the 21-SPONGE survey is limited both in the small number of sightlines used as well as sources being located vast distances from each other on the plane-of-sky, yielding an average of gas that may not physically be interacting or connected.

In this work, we make use of recent observations of Galactic absorption and emission in the direction of the Magellanic system ($-45^{\circ} < b < -25^{\circ}$, $270^{\circ} < l < 300^{\circ}$). The gas of the Milky Way in this direction forms two distinct filamentary structures overlapping each other (\citealp{Nguyen+2024}; Lynn et al., in preparation). The \HI\ survey in this region of the sky comes from the pilot phases of the Galactic Australian Square Kilometre Array Pathfinder (GASKAP) survey, observing the region with a spatial resolution of 30 arcsec \citep{Dickey+2013}. Using the absorption pipeline for GASKAP described in \cite{Dempsey+2022}, the absorption survey searched in an unbiased manner across the full plane-of-sky survey range, producing a large number of 462 detections of Galactic \HI\ absorption.

While the individual detections sightlines are used for studying the physical and spatial properties of cold gas in the cirrus gas of the Milky Way (\citealp{Nguyen+2024}; Lynn et al., in preparation), we instead focus on utilising stacking both the detections sightlines, as well as the non-detections of Galactic \HI\ absorption. Through stacking these two samples independently, we probe the global properties of the gas that lies beyond the sensitivity limits of individual spectra through GASKAP's current operational sensitivity, revealing the higher temperature phases of \HI gas in this region of the sky. The large number of the non-detections also allows for stacking sources based on their spatial distribution to investigate differences across the sky, a technique that was not possible previously.

In this paper, we present analysis of the optical depth and spin temperature of \HI\ gas in the Milky Way towards the Magellanic system and the spatial distribution of these properties through stacking the detection and non-detection sources from the GASKAP Pilot Phase II observations \citep{Nguyen+2024}.
The paper is organized as follows. Section~\ref{sec: Observations} presents the data used in this work. In Section~\ref{sec: Stacking Considerations}, we provide a theoretical background on stacking lines-of-sights with multiple Gaussian components and the consequences for extracted parameters. Section \ref{sec: Stacking Model-Subtracted Detections} investigates stacking the detection spectra after subtracting the already detected cold gas components from \cite{Nguyen+2024}. Section~\ref{sec: Stacking Non-Detections} then describes the stacking analysis of our GASKAP non-detection catalogue using (i) the whole sample, and (ii) spatially distinct bins based on their proximity to known locations of cold gas. A discussion and summary are provided in Sections~\ref{sec: Discussion} and \ref{sec: Summary}.

\section{Data}
\label{sec: Observations}

\subsection{Observations}
\label{subsec: Observations}

We made use of observations from the GASKAP-HI Pilot Phase II survey using the Australian Square Kilometre Array Pathfinder (ASKAP) \citep{Dickey+2013}. 
The observations were taken from 2021 October (SBID 33047) to 2022 April (SBID 38845) with each field gaining 10 h of exposure.
The Pilot Phase II observations comprise ten 25 square-degree adjacent fields of view covering the entire Large Magellanic Cloud, Small Magellanic Cloud, as well as the whole Magellanic Bridge. 
Put together, the fields form a large mosaic from which we have analysed the background sources seen in absorption against the Milky Way foreground gas that was mapped at local velocities.
\subsection{Source catalogues}
\subsubsection{Detections}
\begin{figure}
	\includegraphics[width=\linewidth]{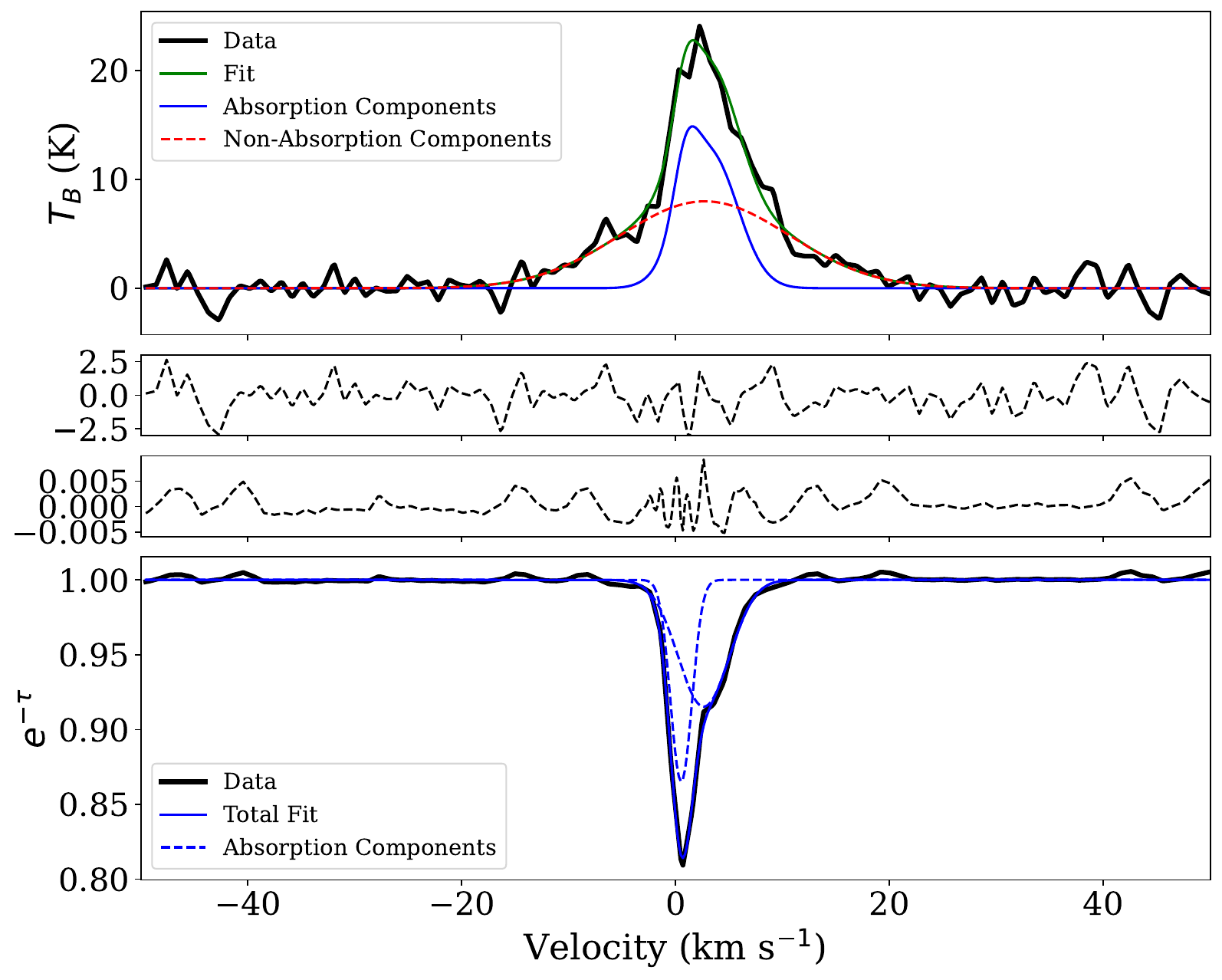}
 	\includegraphics[width=\linewidth]{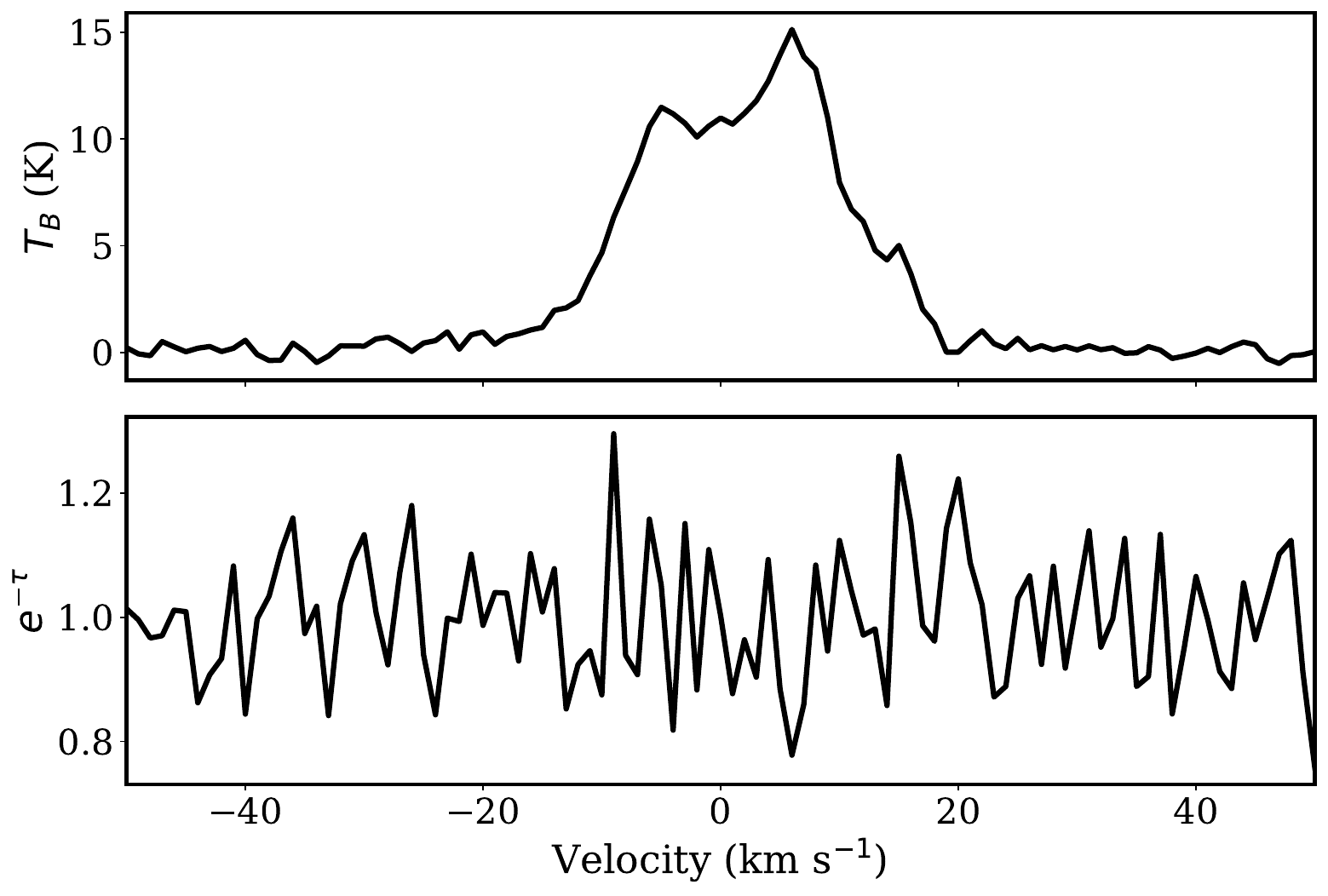}
    \caption{Top panels: Brightness temperature $T_B$ in units of kelvin of the absorption source J040848-750719 from the detection catalogue (top), and its corresponding absorption profile $e^{-\tau}$ (bottom). The coloured lines show the Gaussian decomposition as documented in \citet{Nguyen+2024}, with the fit of the Gaussian components found in absorption (dashed blue lines) along with their total contribution in emission (solid blue), an additional component added in emission tracing warm gas not found in absorption (dashed red) and the total fit of the emission profile (green). Bottom panels: Brightness temperature $T_B$ (top) and its corresponding absorption profile (bottom) of a randomly selected line-of-sight with no detected Galactic \HI\ absorption.}
    \label{fig: Detection Example}
\end{figure}

\citet{Nguyen+2024} utilised the GASKAP absorption pipeline developed by \citet{Dempsey+2022} to process 2714 sources of \HI\ absorption and their spectra. 
Twelve of these absorption spectra were saturated and removed from the sample. The remaining 2702 sources selected had a background source flux, $S_{\rm cont}$, greater than 15 mJy.
Of these spectra, 462 were deemed to contain detections of \HI\ absorption from gas pertaining to the Milky Way, defined as having a point above $3\sigma$ and a neighbouring channel above $2.8\sigma$. \cite{Nguyen+2024} used Gaussian decomposition to then analyse the spin temperature and optical depth properties of the cold gas in these detection sightlines.

Relevant for this work, Fig. \ref{fig: Detection Example} (top panels) displays one of these absorption detections (bottom) against the radio source J040848-750719 and its associated emission spectra interpolated around the source (top). It is noteworthy that there are multiple peaks seen in absorption against this source. The blue and red lines show the modelled contribution from the cold and warm gas obtained following the methodology prescribed in \cite{Heiles_Troland+2003}.

The mean rms noise of the detection sample is approximately $\sigma_{\tau} = 4.2\times10^{-2}$ and a median of around $3.1\times10^{-2}$, around 15 times higher than the noise level of individual channels in more sensitive observations like 21-SPONGE \citep{Murray+2014,Murray+2018}. However, the main advantage with the GASKAP survey is the order of magnitude increase in the number of sightlines available to stack.

\subsubsection{Non-detections}

\begin{table}
\caption{Sample of GASKAP Pilot Phase II non-detection parameters.}
\centering
\begin{tabular*}{\columnwidth}{l rrrr}
\hline\hline
Source &     l &     b &  $\sigma_{\tau}$ $^{a}$ & $v_{\rm peak}$ $^{b}$ \\
 &     ($^{\circ}$) &     ($^{\circ}$) &  ($\times10^{-3}$) & (\kms) \\
\hline
 J055932$-$634614 & 273.2 & $-$29.8 &  244.5 &         2 \\
 J044109$-$631421 & 273.9 & $-$38.5 &  158.0 &        12 \\
 J053738$-$624208 & 272.0 & $-$32.3 &   93.3 &         6 \\
 J060128$-$624704 & 272.0 & $-$29.6 &  204.9 &        $-$1 \\
 J053234$-$631640 & 272.7 & $-$32.8 &  149.0 &         5 \\
 J060038$-$674747 & 277.8 & $-$29.7 &  154.8 &        $-$5 \\
 J003524$-$732222 & 304.5 & $-$43.7 &  116.5 &         0 \\
 J044926$-$691203 & 280.7 & $-$35.9 &   88.5 &         3 \\
 J050107$-$691737 & 280.5 & $-$34.9 &  190.4 &         4 \\
 J013229$-$734122 & 299.0 & $-$43.1 &  238.4 &         1 \\
\hline
\end{tabular*}
\raggedright
\textbf{Notes:} \\
$^{a}$: Mean rms noise in optical depth over all velocity channels. \\
$^{b}$: Line-of-sight velocity of the peak in brightness temperature.
\label{table: List of Sources}
\end{table}

In this work, we also used the remaining 2240 sources where there were no detections of \HI\ absorption with a signal-to-noise greater than 2.8 in multiple neighbouring channels. Hereafter, we will refer to them as the non-detection sample.
An example of an absorption-emission pair selected randomly from the non-detection sample is shown in Fig. \ref{fig: Detection Example} (bottom panels).
While the sensitivity of the absorption profile is too low to detect any features, the associated emission spectrum shows prominent peaks, one of which is relatively narrow, that reveal the presence of CNM gas mixed within the warmer phases of the ISM (i.e., the UNM and the WNM).

Table \ref{table: List of Sources} lists the source name, Galactic coordinates $l$ and $b$, average rms noise $\sigma_{\tau}$ over all velocity channels, and velocity at the peak brightness temperature $v_{\rm peak}$ of ten sample entries of the 2240 non-detection source catalogue.
The mean rms noise of the non-detections is $\sigma_{\tau} = 0.18$ with a median rms noise of approximately $0.14$. The noise level of the GASKAP absorption sightlines is several orders of magnitude higher than that of 21-SPONGE \citep{Murray+2018}. However, similar to the detection sample, this noise level increase is countered by having two orders of magnitude more sources than that used in the stacking analysis in \cite{Murray+2014,Murray+2018}.

\subsubsection{Flux versus sensitivity}
\label{subsec: Flux vs Sensitivity}

\begin{figure}
	\includegraphics[width=\columnwidth]{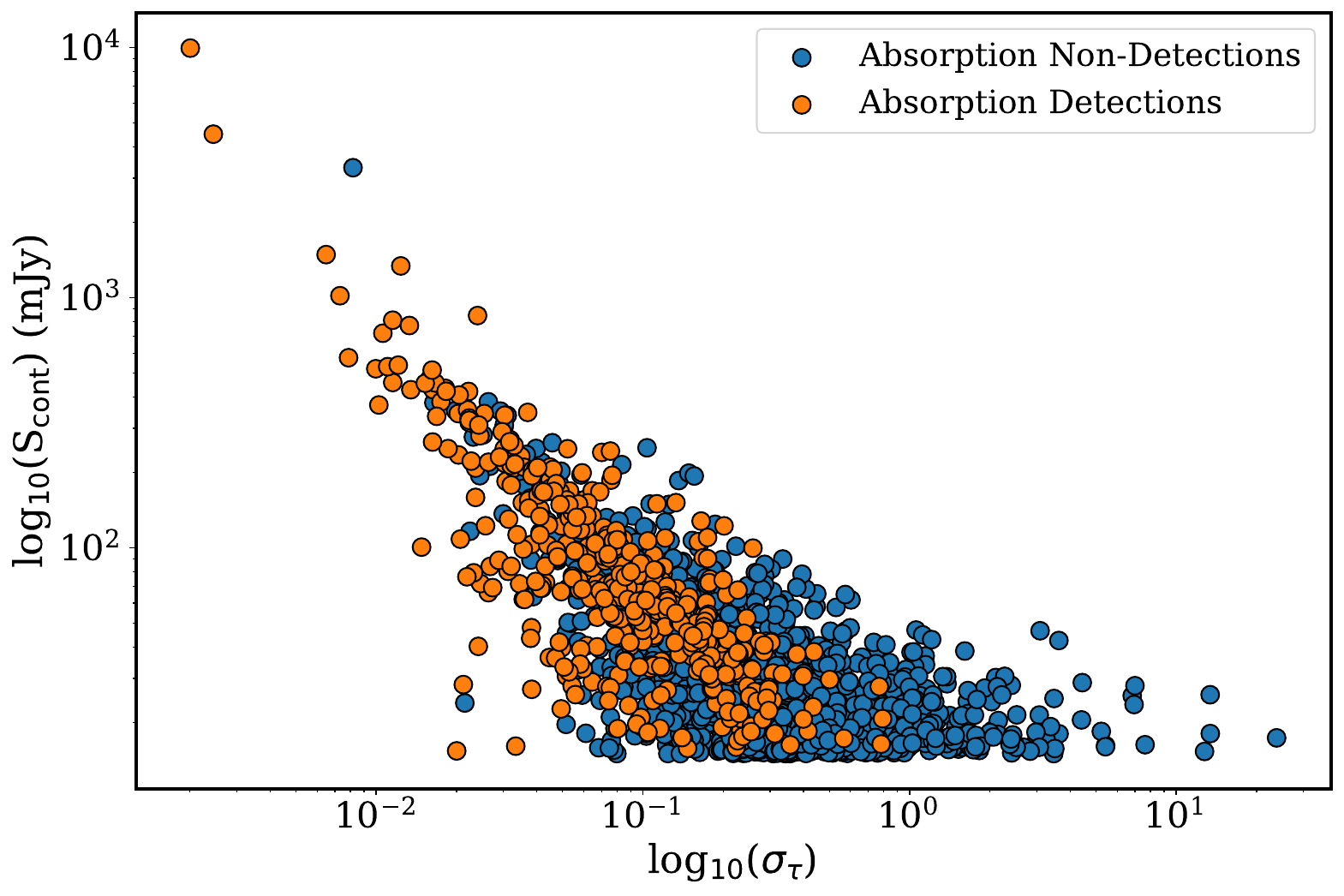}
    \caption{Scatter plot of the background source flux, $S_{\rm cont}$, versus the optical depth noise, $\sigma_{\tau}$, for each source of the 2240 non-detections (blue) and the 462 detections (orange) of Milky Way \HI absorption. All sources, regardless of detection, have a background flux higher than the cut-off of $15$\,mJy \citep{Dempsey+2022}.}
    \label{fig: Flux vs Sigma Tau Comparison}
\end{figure}

A comparison between the two data sets of detections and non-detections is shown in Fig. \ref{fig: Flux vs Sigma Tau Comparison}, comparing their background source flux and the mean rms noise of their absorption spectra. Detections and non-detections are shown in orange and blue, respectively. We note that the detections typically have higher background source fluxes, where there is a higher probability of the line-of-sight (LOS) producing a detection of absorbing gas. Most of the detections are within the same parameter space as the non-detections. However when the noise level is greater than $\sigma_{\tau}\sim 3\times10^{-1}$, the spectra are too noisy for any gas regardless of its opacity to produce a signal beyond the $\sim 3\sigma$ signal-to-noise requirement of a detection.

Since the non-detections spectra have similar background source fluxes to most of the detections, there is nothing intrinsically different about these spectra other than the higher noise levels. Thus, while the non-detection sources did not meet the criteria for being used in regular absorption analysis \citep{Dempsey+2022,Nguyen+2024}, they should nonetheless still contain astrophysical signals buried within their noise.

\subsection{Emission data}

\begin{figure*}
	\includegraphics[width=\linewidth]{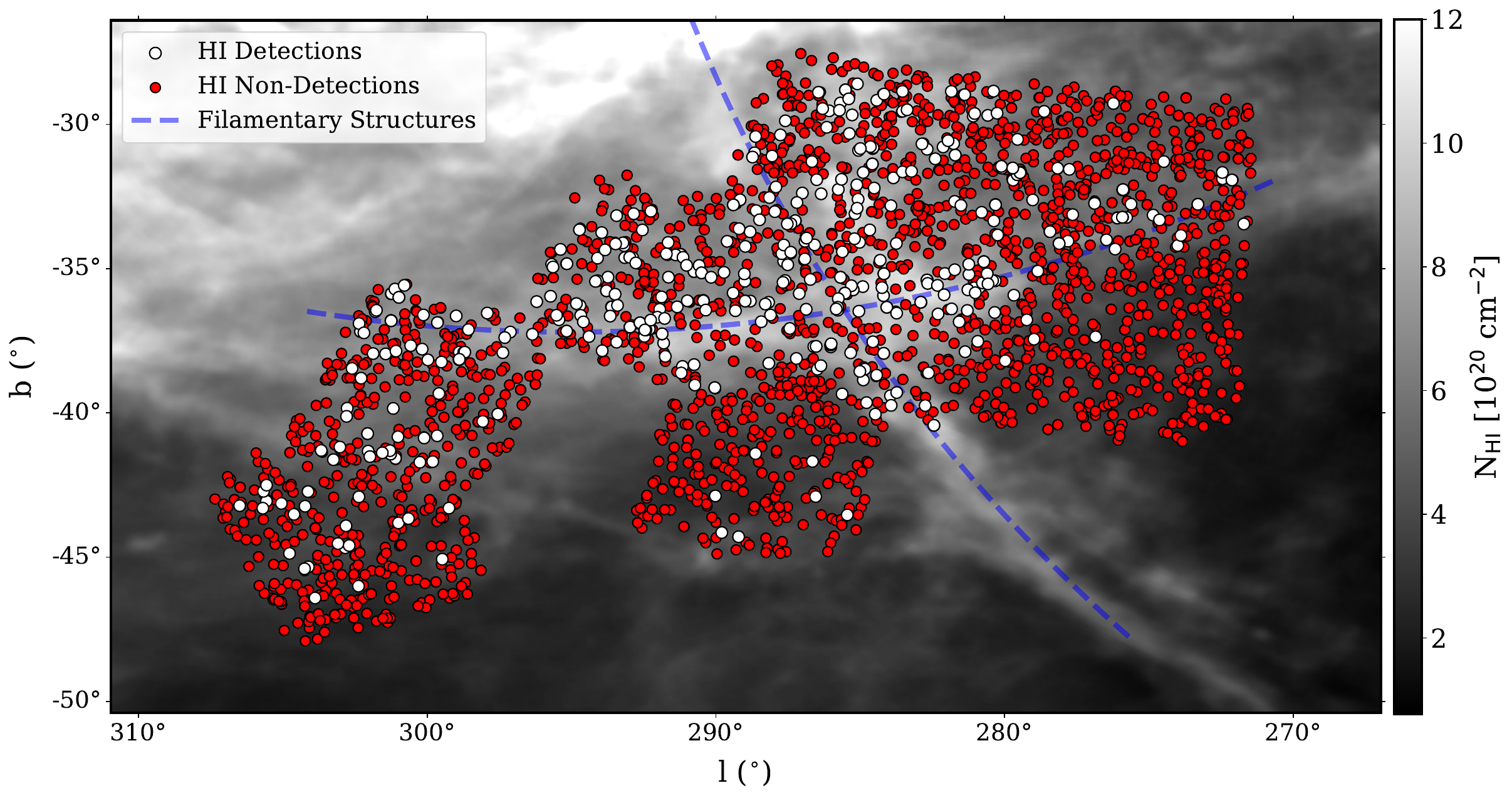}
    \caption{Spatial distribution of background sources from the GASKAP-HI absorption survey overlaid on a grey-scale map of the column density of \HI\ gas computed in the optically thin limit from the GASS survey \citep{Mcclure-Griffiths+2009,Kalberla+2010,Kalberla_Haud+2015}. Detections of cold gas in the velocity range $-50 < v < 50$\,\kms\ are annotated with white circles. Sightlines with non-detections of cold gas in absorption are shown with red circles. The blue dashed lines indicate the arcs of the vertical and horizontal \HI filaments in this region of the Milky Way (Lynn et al., in preparation).}
    \label{fig: Field of View of Non-Detection Sources}
\end{figure*}

To pair these absorption sightlines with emission data, we made use of the HI emission data provided by the GASKAP Pilot surveys as mentioned in Section \ref{subsec: Observations}. The emission data have a beamsize of 30 arcsec and native spectral resolution of 0.24 \kms. For noise reduction, the emission and absorption spectra were binned spectrally by a factor of 4 to a spectral resolution of $\sim$ 1 \kms. As the emission sightline located at the position of HI absorption is affected by the absorbing feature, emission spectra from nearby sightlines on the plane-of-sky were used instead, assuming that the gas within this distance is similar. These emission sightlines also have to be far away enough from each other that beam smearing does not affect their average. For this purpose, we used a method similar to that of \cite{Nguyen+2024}. For each absorption sightline, approximately 20 emission sightlines were selected around the absorption feature, within an annulus of inner-radius 1 beam-radius and outer-radius 2.5 beam-radii, ensuring each is at least 1 beam-radius from other sightlines. These emission sightlines are then averaged, weighted by the inverse of their plane-of-sky distance from the absorption sightline at the centre.

Fig.  \ref{fig: Field of View of Non-Detection Sources} shows the locations of the absorption sightlines, including both detections and non-detections at Milky Way velocities in the velocity range $-50 < v < 50$\,\kms\ overlaid onto the column density map of the region taken from the Galactic All-Sky survey computed in the optically thin limit \citep[GASS, surveyed with the Parkes telescope with a beam size of 16$^{\prime}$ and 1\kms\ spectral resolution;][]{Mcclure-Griffiths+2009,Kalberla+2010,Kalberla_Haud+2015}. The column density map highlights the vertical and horizontal \HI\ filamentary structures in the region, defined by high \HI\ column density and high brightness temperature (Lynn et al., in preparation).

\section{Stacking Considerations}
\label{sec: Stacking Considerations}

When stacking multiple spectra to produce an approximate average spectra over a region, one needs to take into consideration the effects of averaging multiple Gaussian components that are potentially misaligned in velocity.
In this section, we explore, through use of the GASKAP detection sample, the biases introduced by mixing complex multiphase emission and absorption spectra through stacking.

\subsection{Velocity alignment}
\label{subsec: Velocity Alignment}
When stacking spectra, the first thing we must consider is how to account for effects like Galactic rotation, or bulk motions of gas that could displace components in line-of-sight velocity. For this work we explore shifting the emission and absorption spectra such that the peak brightness temperature is centred on 0 \kms. The peak in emission typically traces cold gas and so shifting to centre this ensures that at least the bulk of \HI cold gas is centred on the same velocity when stacking. With sightlines containing single components, this method of shifting perfectly aligns all components to produce a true average spectrum. We note that in reality some minor deviations are present between the peak in emission and absorption that are typically on the order of $1-2$ \kms, as observed in the GASKAP detections \citep{Nguyen+2024}. These deviations are due to a combination of the difference in noise-levels between the emission and absorption spectra as well as the emission and absorption spectra probing slightly different gas as they are not along the same sightline. These effects cause some uncertainty in the central velocity of the peak of absorption and its alignment with the peak in emission. However, the $1-2$ \kms\ deviation is negligible compared to the $\sim10-15$ \kms\ variation in central velocities of cold components in this region of the sky \citep{Nguyen+2024}.

Other methods in the past have utilised shifting by the first velocity moment \citep{Murray+2014,Murray+2018} producing similar results with single components as this simplifies to centring by the peak. However, the two shifting methods deviate when we consider sightlines with multiple components. As shown in the example detection spectra of Fig.  \ref{fig: Detection Example}, the most sensitive detection sources showcase multiple components: a narrow primary component aligned well to the peak in the respective emission profile and a slightly broader secondary component offset in velocity. Therefore, velocity shifting needs to be able to accommodate for the existence of multiple clouds along a sightline. In a multiphase medium with multiple components per sightline, none of the components along a line-of-sight will be perfectly centred to 0 \kms\ if shifted by the first velocity moment. When stacking these partially shifted spectra, the key features of colder and warmer phases will be heavily mixed together.

Since in a general stacking case, we are unsure of the number of components in each sightline prior to stacking, shifting by the first velocity moment will make it difficult to separate out the phases of \HI through Gaussian decomposition.
If we shift instead by the peak in emission, we avoid these problems, as we standardise the individual sightlines such that the bulk of cold gas (traced by the peak in emission and henceforth referred to as the "primary component") is centred to the same velocity. While additional components at other local maxima will be mixed together since their central velocities are not accounted for (producing what will be referred to as the "secondary component"), shifting in this manner allows for easier separation of the phases of \HI with a major part of the cold phase aligned in velocity at 0 \kms\ prior to stacking.

\subsection{Stacked spectrum}
\label{sec: Stacked Spectrum Detections}

The GASKAP detection sample provides a unique opportunity to test how stacking affects measured properties, given that the sightlines are all in the same plane-of-sky location, and sightlines contain multiple cold components \citep{Nguyen+2024}.

Our stacking analysis follows the procedure described in \citet{Murray+2014}, where they formed a stack of 19 absorption spectra from the 21-SPONGE survey, after subtracting Gaussian fits for cold gas, to investigate properties of the WNM. For this section, however, we are using the un-subtracted detection spectra as observed, containing all the phase information of the ISM to compare the stack with the average properties of the individual sightlines.

The stacked spectrum was obtained by taking the average of the sample weighted by the noise of each absorption spectrum, with weights $\tau_{\rm res}/\sigma_{\tau}^{2}$ \citep{Treister+2011} to favour spectra with lower noise. The weights simplify to $1/\sigma_{\tau}$ since $\sigma_{\tau} \propto T_{\rm B}$ and for constant temperature, $\tau_{\rm res} \propto T_{\rm B}$ \citep{Murray+2014}. The same weights were applied in the stack of the emission spectra. Fig.  \ref{fig: Detection Stack Gaussian Decomposition} shows the stacked emission and absorption spectra (top and bottom panel, respectively). The absorption stack produces a two-component profile; a deep, narrow, primary component, and a broad secondary component. 

\subsection{Gaussian decomposition}
\label{subsec: Gaussian Decomposition}

\begin{figure}
	\includegraphics[width=\columnwidth]{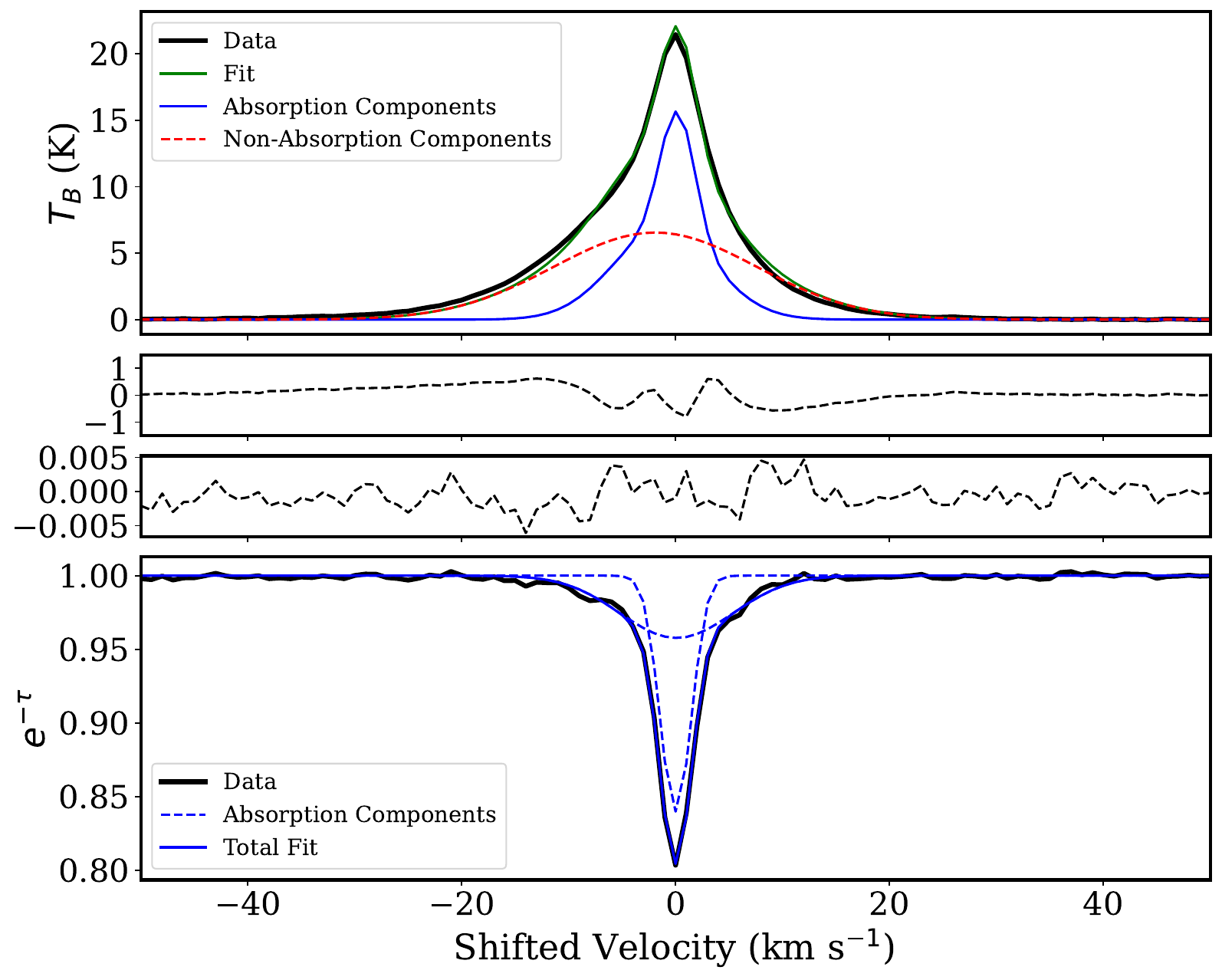}
    \caption{Top: Weighted-mean emission profile of the stacked GASKAP detection spectra.
    The original stacked spectrum is shown in black and the model of the joint Gaussian decomposition in green. The model is made by the addition of the emission produced by the narrow and broad components (shown in blue), and an additional broad component required to complete the fit (shown by the red dashed line). Middle: Residual profiles of the emission (middle top) and absorption (middle bottom) fits. Bottom: Weighted-mean absorption profile of the stacked GASKAP detections. The original stacked spectrum is shown in black. The narrow and broad components' absorption fits are shown by the blue dashed lines, with their total contribution to the absorption profile shown in the solid blue line.}
    \label{fig: Detection Stack Gaussian Decomposition}
\end{figure}

To properly Gaussian decompose the stacked spectra, we implemented the method prescribed in \citet{Heiles_Troland+2003} to model their spin temperatures individually. Two Gaussian components were fitted to the absorption stacked spectrum. The fitting parameters were then used to model the amplitude of their emission counterparts, allowing for a $\pm1$ \kms\ deviation in their central velocities and 10 per cent deviation in their widths. An additional broader Gaussian component was added to fit the signal in emission to ensure the column density of the fit was not lower than that of the original data from GASKAP, and that the correction factor, $R=N_{\HI}/N^*_{\HI}$, was greater than one, where $*$ denotes the column density computed in the optically thin regime. The additional component in emission represents gas from the WNM that is not able to be detected in absorption due to the insufficient optical depth sensitivity of our stack. The resulting decomposition is displayed in Fig.  \ref{fig: Detection Stack Gaussian Decomposition}. It should be noted that a potential bias in adding only a single emission-only component is that it under-represents the spin temperature of the absorption feature. Adding additional components to fit the same data would lead to them being narrower in width and shallower in amplitude, allowing the absorption components to more closely fill the emission at the peak. This higher amplitude in brightness temperature would lead to a higher recorded spin temperature.

When solving the radiative transfer equation of the line, we accounted for each possible permutation of the broader emission component being in either the foreground or background relative to the colder components seen in absorption along the LOS. For the fraction of warm gas in the background (and hence absorbed by the cold gas infront of it), we used three values: 0, 0.5, and 1. Averaging these solutions to the radiative transfer yielded the mean spin temperatures. The primary and secondary component of the total stacked spectrum using all 462 detection sources have a mean spin temperature of $T_s=68\pm$7 K and $159\pm58$ K, peak optical depth of $\tau_{\mathrm {peak}} = 0.160\pm0.033$ and $0.040\pm0.007$, and full-width half-maximum (FWHM) of $3.4\pm1.4$ \kms\, and $12.5\pm1.1$ \kms\,, respectively. The uncertainties are produced through a method of bootstrapping, where new samples of 462 detection spectra were sampled from the original 462 sightlines with replacement. Repeating the stacking and fitting process of the new sample, and repeating this $10^{5}$ times, the standard deviations of the above measurements of $T_{s}$, $\tau_{\mathrm {peak}}$, and FWHM were computed. 

\subsection{Comparison of stack versus original sample}
\label{subsec: Comparison of Stack vs Original Sample}

\begin{figure}
	\includegraphics[width=\columnwidth]{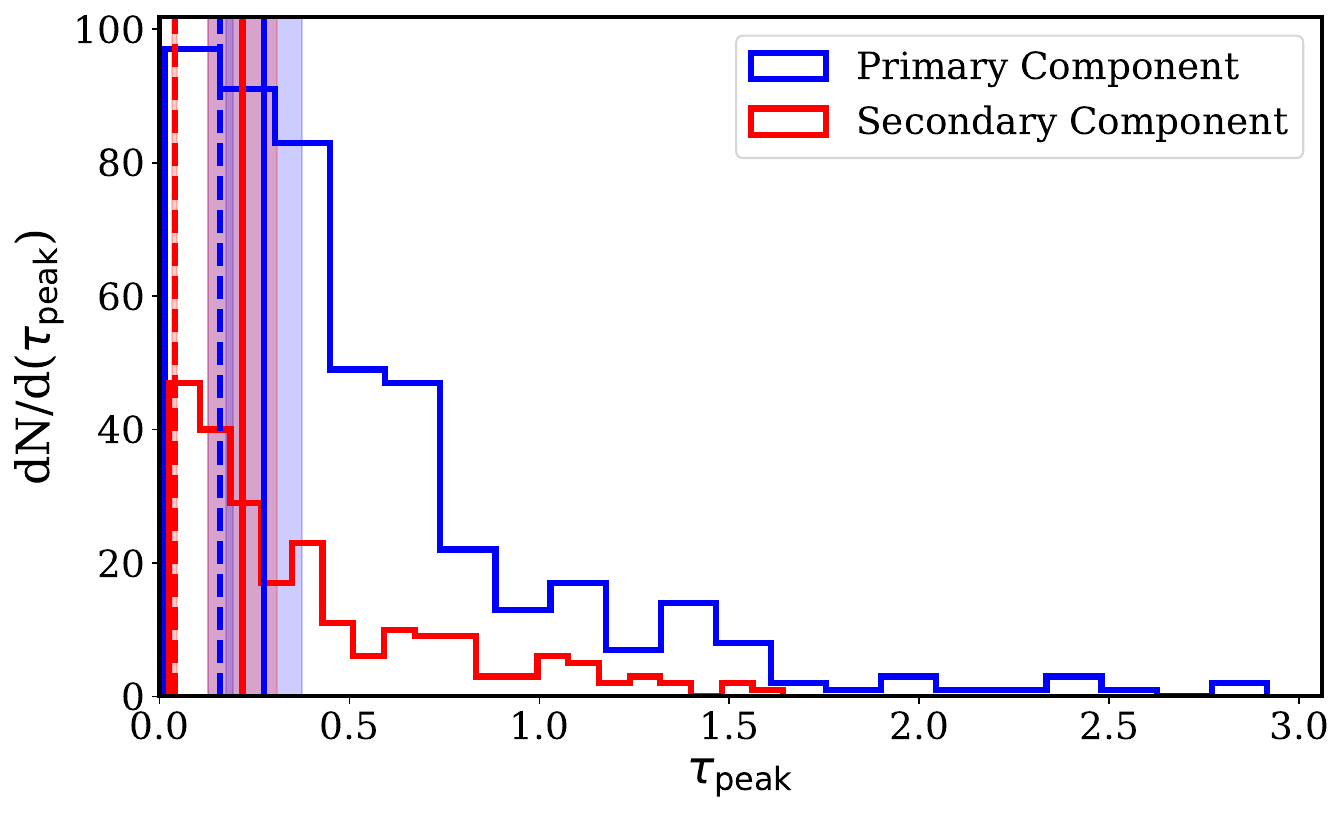}
    \includegraphics[width=\columnwidth]{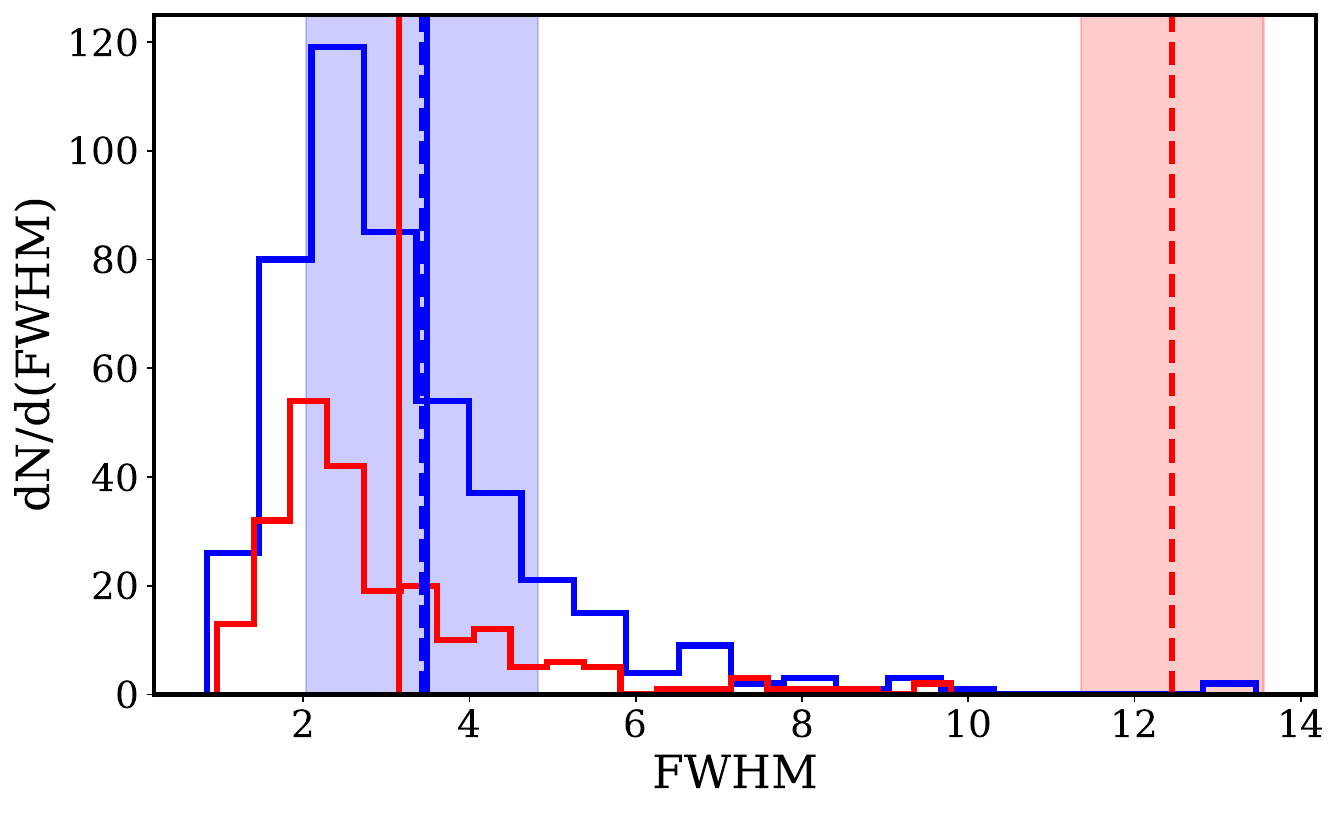}
    \includegraphics[width=\columnwidth]{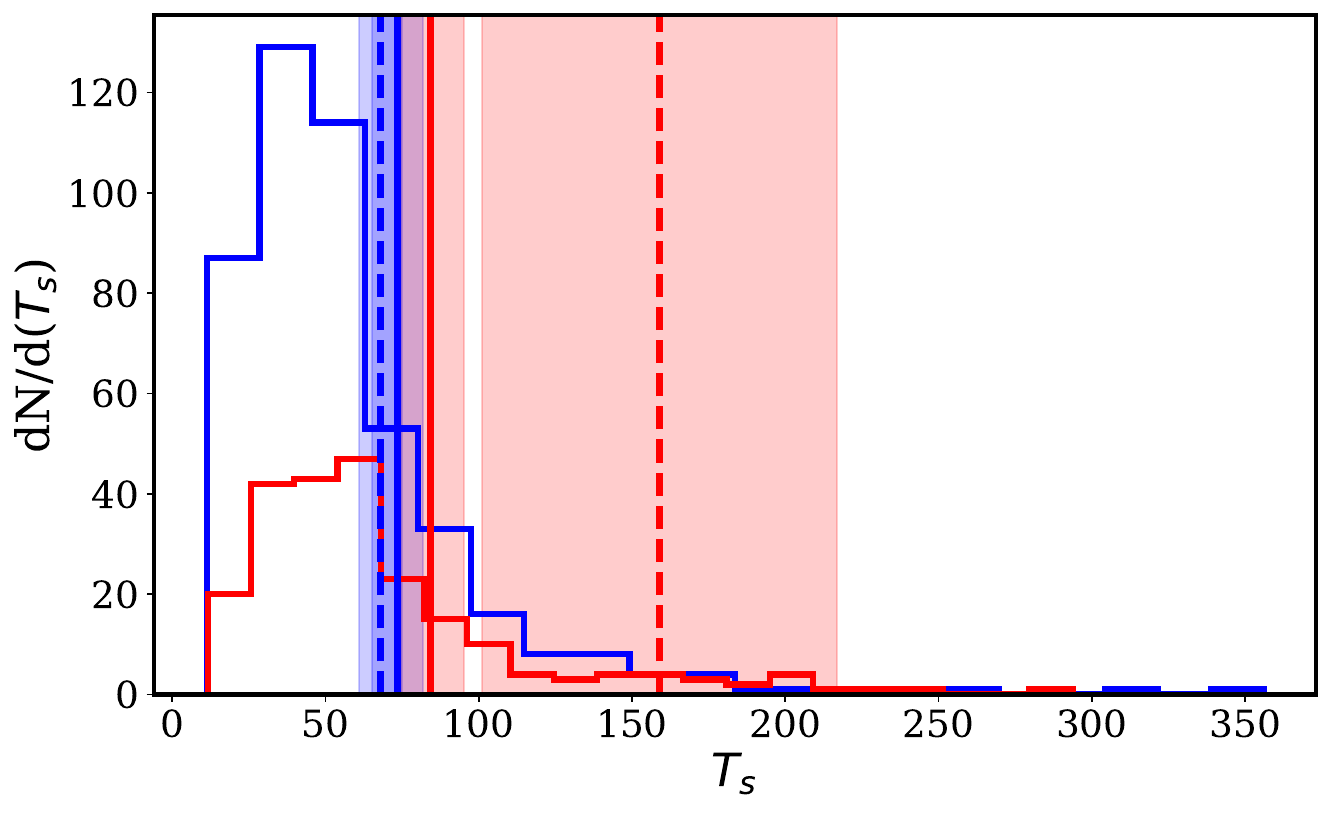}
    \caption{Distributions of the peak optical depth (top), FWHM (middle), and spin temperature (bottom) of individual absorption components found by \citet{Nguyen+2024}. The components are separated into primary components (blue), which are aligned with the peak in their corresponding emission spectra, and secondary components (red), containing additional components offset from the emission peak. The solid vertical lines represent the mean of these distributions in their respective colours. The dotted lines highlight the values obtained from the detection stack in Fig. \ref{fig: Detection Stack Gaussian Decomposition} with the shaded regions representing the uncertainties of both the mean and observed stacked values.}
    \label{fig: Detection Individual Sightlines vs Observed}
\end{figure}

To examine the effect of stacking spectra on the measurements properties of spin temperature, optical depth, and FWHM, we can compare these values to that of the individual detection sightlines, using the properties extracted by \cite{Nguyen+2024}. Following the discussion in Section \ref{subsec: Velocity Alignment}, absorption detection components located closest spectrally to the peak in emission in each detection sightline are considered primary components. The remaining absorption components are collected as the secondary components in each spectra. Fig.  \ref{fig: Detection Individual Sightlines vs Observed} shows the comparison between the observed peak $\tau$, FWHM, and spin temperatures of both stacked components and the distribution of the individual sightlines' primary and secondary components. We generate the weighted mean of these individual sightline properties using the same weighting of the inverse of the optical depth noise that produced the stacks. The uncertainty in these means were calculated by adding the uncertainty of the mean and the systematic errors in quadrature. The systematic errors were approximated as the mean of the uncertainties of individual sightline values in $\tau_{\mathrm {peak}}$, FWHM and $T_{\mathrm s}$. When considering the overlap of the errors in the stacked values and that of the mean individual values, we find that the primary component's optical depth and FWHM are both accurate representations of an average value. However when it comes to the secondary component, the observed peak optical depths lies below the mean of the individual sightline peak $\tau$ measurements that produced them. Conversely, the FWHM measurements observed in the secondary component's stack lie above the mean FWHM of the individual sightline components.

The reason for this discrepancy between the secondary component's stacked values compared to the mean properties of the individual sightlines comes from the velocity offset of components. Since the components are slightly offset in velocity from each other, the width of the resulting stacked Gaussian will be wider than the average of the individual Gaussians. Due to the total flux of all the components being conserved in the average stack, if the width of the Gaussians is increased, to compensate the amplitude of the resulting stacked Gaussian must diminish. Therefore, regardless of the severity of the range of central velocities of secondary components in individual sightlines, the peaks will always be lower limits and their widths always greater than the average of the individual components themselves.

The primary components on the other hand, aligned with the peak in emission, are shifted correctly to roughly $0$ \kms\ with some minor deviation due to possible slight offset between absorption and emission peaks in velocity (as mentioned in Section \ref{subsec: Velocity Alignment}). As such the primary component's lack of a noticeable velocity offset leads to the stacked result representing a relatively accurate average of the peak optical depth and FWHM.

When calculating the spin temperatures of the components, the primary component's stack and mean values are well aligned along with their uncertainty envelopes. The secondary component's values are also similar. While their uncertainty envelopes do not perfectly align, it should be noted that due to the potential biases in spin temperature by only including one emission component (as mentioned in Section \ref{sec: Stacked Spectrum Detections}), it is likely we are undersampling the error in the observed stacked temperature. Therefore we conclude that both the primary and secondary components' spin temperatures are well aligned with the means of the original distributions. This is due to the fact that since the velocity offset of components is equal in both emission and absorption, the peaks of emissions and absorption are affected in a similar way. Since the spin temperature is found through the division between peaks in brightness temperature and optical depth, the velocity offset effect is cancelled out, resulting in a spin temperature that is well correlated to the true average temperature of the individual sightlines' components. An exploration into this velocity offset effect is explored in Appendix \ref{sec: Toy Model Theoretical Considerations} where toy models are used instead of real detection data, yielding similar results.

\subsection{Theoretical implications}
\label{subsec: Theoretical Implications}

From stacking the detection sample and comparing the result with the individual sightline properties (as well as through the use of toy models in Appendix \ref{sec: Toy Model Theoretical Considerations}), we find several effects to be considered.

The uncertainty of the true peaks and widths of the stacked components means we should be cautious when calculating secondary information extracted from emission-absorption pair stacks that use these values such as the FWHM, corresponding turbulent Mach numbers and mass fractions, as they become unreliable.

This is especially true for extragalactic observations, where individual sightlines' beamsizes pass through much larger portions of a galaxy than Galactic observations. The large amount of extragalactic ISM material mixed together at various velocities will broaden the Gaussian components in a similar manner to that of the secondary component explored in this section, with individual sightlines acting as a pseudo-stack of the region. As a result, a similar effect will occur and peak optical depth measurements of extragalactic sources should take this lower-limit effect into account when determining results.

The rest of this work will calculate and report parameters including the peak optical depth, FWHM and spin temperature. However, other quantities such as the turbulent Mach number, column density and mass fraction will not be discussed as they develop large uncertainties, as mentioned above.

\section{Stacking Model-Subtracted Detections}
\label{sec: Stacking Model-Subtracted Detections}

In this section, we investigate the existence of higher temperature HI gas in the region of the GASKAP HI Pilot survey by stacking the GASKAP detection spectra after subtracting the cold gas already detected from \cite{Nguyen+2024}.

\subsection{Model subtraction}
\label{subsec: Model Subtraction}

Higher temperature phases of HI tend to have weaker signals in absorption due to lower optical depths. In order to detect these higher temperatures in absorption, we must first subtract the cold gas already detected from individual sightlines. For each of the 462 detection sightlines, \cite{Nguyen+2024} catalogues the fitted model parameters of the cold gas detected in absorption and emission. The modelled Gaussian absorption and emission components, $\tau_{i}(v)$ and $T_{\mathrm B, i}$ (where $i$ indicates the $i$th component) were subtracted from the respective spectra from which they were detected to produce 462 residual absorption spectra, $\tau_{\mathrm res}(v) = \tau(v) - \Sigma\tau_{i}(v)$, and residual emission spectra $T_{\mathrm B, res}(v) = T_{\mathrm B}(v) - \Sigma T_{\mathrm B, i}(v)$.

\subsection{Stacked spectrum}
\label{subsec: Stacking Subtracted Spectra}

Since we have removed the majority of the signal through subtracting the cold gas components, we no longer have information about the velocity of the remaining underlying absorption hidden beneath the noise in individual detection spectra. We follow the same procedure as in Section \ref{subsec: Velocity Alignment}, relying on the GASKAP emission data to align the spectra before performing the stack to account for Galactic rotation and large scale bulk motions in the region. 

\begin{figure}
	\includegraphics[width=\linewidth]{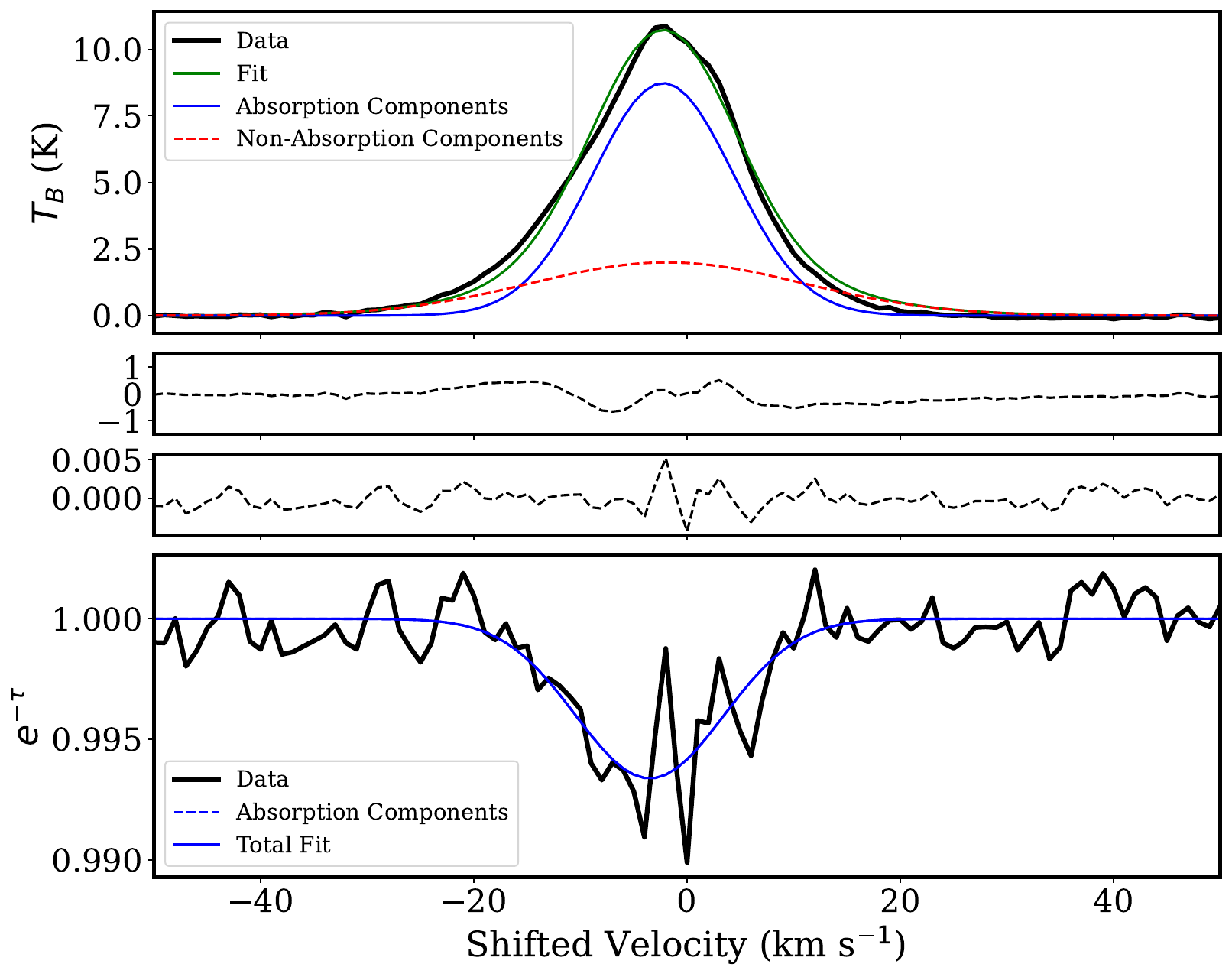}
    \caption{Top: Weighted-mean emission profile of the stacked GASKAP detection spectra after subtracting the cold gas models in individual sightlines from \citet{Nguyen+2024}.
    The original stacked spectrum is shown in black and the model of the joint Gaussian decomposition in green. The model is made by the addition of the emission produced by the absorption component (shown in blue), and an additional broad component required to complete the fit (shown by the red dashed line). Middle: Residual profiles of the emission (middle top) and absorption (middle bottom) fits. Bottom: Weighted-mean absorption profile of the stacked GASKAP non-detections. The original stacked spectrum is shown in black. The absorption fit is shown by the solid blue line.}
    \label{fig: Subtracted Stack Gaussian Decomposition}
\end{figure}

To produce the stacked model-subtracted emission and absorption spectra, we use the same weighting as in Section \ref{sec: Stacked Spectrum Detections} following \cite{Murray+2014}. The resulting stacked emission and absorption spectra are shown in Fig.  \ref{fig: Subtracted Stack Gaussian Decomposition}. The rms noise of the stacked spectrum is $\sigma_{\tau} = 1.1\times10^{-3}$, calculated from taking the standard deviation of 30 empty channels in the post-shifting velocity range $-50 < v < -20$\,\kms. The stacked detection spectra's noise level is almost 40 times smaller than the mean rms noise of the individual detection sightlines ($\sigma_{\tau} = 4.2\times10^{-2}$) and nearly 30 times smaller than their median rms noise ($\sigma_{\tau} = 3.1\times10^{-2}$). The stacked absorption profile has an equivalent width (EW) in the velocity range $-50 < v < 50$\,\kms\ of EW $= \int_{-50}^{50} {(1 - e^{-\tau}) \ dv} = 0.12 $\kms \citep{Murray+2014}.

\begin{figure}
	\includegraphics[width=\columnwidth]{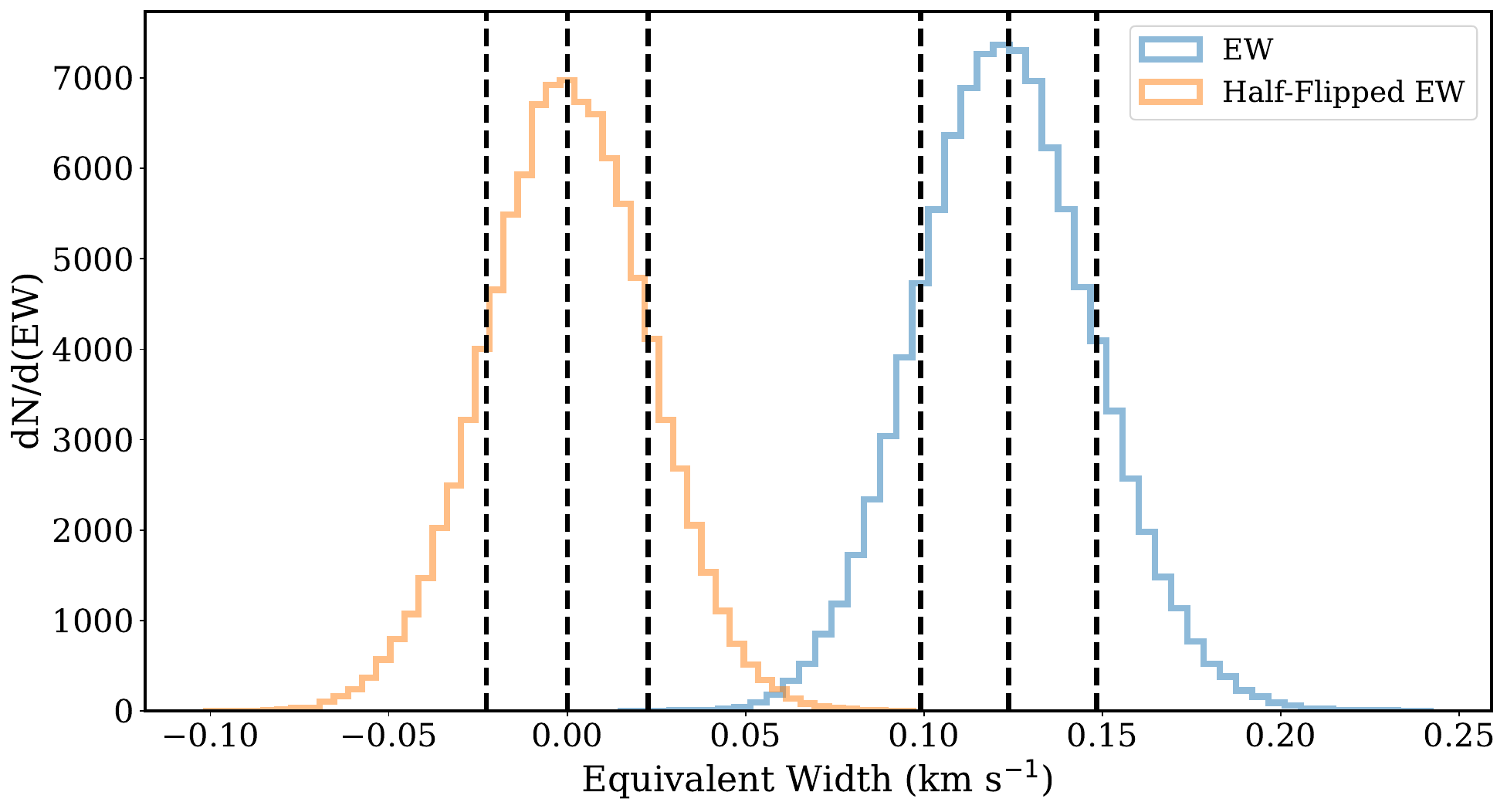}
    \caption{Histograms of the EW distribution from $10^{5}$ trials of bootstrapping, randomly sampling 462 spectra from the model-subtracted detection sample derived in Section \ref{subsec: Model Subtraction}, with possibility of replacement. The blue histogram displays the distribution of EW values from the trials, with a mean of $\langle $EW$ \rangle = \langle \int_{-50}^{50} {(1 - e^{-\tau}) \ dv} \rangle = 0.12 \pm 0.02$\kms\ ($68$ per cent confidence). The orange histogram displays the same random trials except with half of the spectra inverted (multiplied by $-1$) before stacking to test for contamination from outlying spectra. The mean of the orange distribution is $0.00 \pm 0.02$\kms\ ($68$ per cent confidence). The vertical dotted line crossing the middle of the blue histogram represents the EW of the original stack (EW$=0.12$ \kms). The dotted line along the middle of the orange histogram represents the expected mean of $0$ \kms\ for the half-inverted stack EW. The additional dashed lines mark the $68$ per cent confidence interval for the two distributions.}
    \label{fig: EW Distribution Subtracted}
\end{figure}

To ensure that our stack is not plagued by any outlying spectra and the detected component of the stack is not produced by only a few dominant narrow profiles in the average, we conducted two bootstrapping Monte Carlo simulation experiments \citep{Wall+2003,Murray+2014} to test for variations in the stack's EW. 
First, we repeated the stacking analysis on a new sample of 462 sources randomly chosen from our original 462 sightlines with replacement. Repeating this analysis $10^{5}$ times, we computed the EW for each trial. The distribution of these EW values is shown in Fig. \ref{fig: EW Distribution Subtracted}. The distribution is Gaussian, with a mean value of $\langle EW \rangle = 0.12 \pm 0.02$ \kms\ ($68$ per cent confidence), suggesting that the EW computed from the original sample is consistent with being drawn from a parent population of spectra with a comparable mean and is not influenced by outlying spectra. Secondly, we produced a similar EW distribution where in each trial we inverted (multiplied by $-1$) half of our randomly selected sample. The distribution, also shown in Fig. \ref{fig: EW Distribution Subtracted}, is centred around $0$\kms, consistent with no signal. The result verifies that our stack does not produce a false signal.

We used the method described in Section \ref{sec: Stacked Spectrum Detections} to decompose the stacked spectrum and determine the properties of the remaining gas. The stacked spectrum produces a single wide component in absorption, with a peak optical depth of $\tau_{\mathrm {peak}} = (6.6\pm0.5)\times10^{-3}$, FWHM of $16.3\pm1.4$ \kms, and $T_{s} = 1320\pm263$ K. The uncertainties in these values were evaluated by repeating the same fitting procedure to each of the $10^{5}$ trials of the bootstrapping sample used earlier. These uncertainties are larger than those from the variation of the fraction of background warm gas in the radiative transfer solution.

\section{Stacking Non-Detections}
\label{sec: Stacking Non-Detections}

Stacking the detection spectra is useful for investigating higher temperature gas in sightlines where we know cold gas already exists. However, the GASKAP survey has the unique opportunity to use 2240 non-detection spectra in the same region on the sky that likely also contains information about the multiphase ISM buried within their noise. In this section, we investigate stacking the GASKAP non-detection spectra to reveal more information about cold gas in this region of the Milky Way.

\subsection{Velocity alignment}
\label{subsec: Velocity Alignment Non-Detections}

Since there is no information about the velocity of the underlying absorption line hidden beneath the noise in individual non-detection spectra, we rely on the peak in the GASKAP emission data to align them before performing the stack to account for Galactic rotation and large scale bulk motions in the region (as explained in Section \ref{subsec: Velocity Alignment}).

\begin{figure}
	\includegraphics[width=\linewidth]{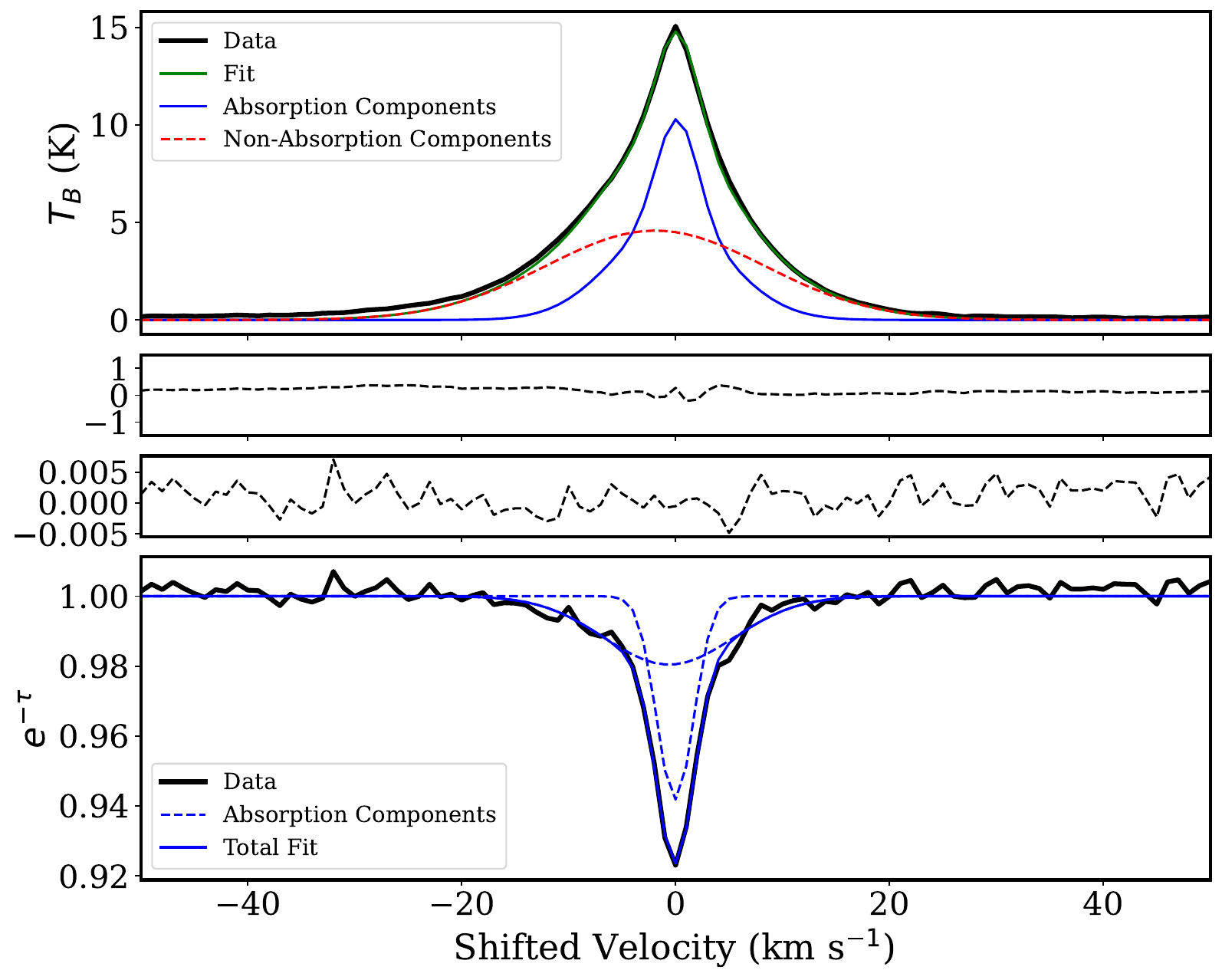}
    \caption{Top: Weighted-mean emission profile of the stacked GASKAP non-detections.
    The original stacked spectrum is shown in black and the model of the joint Gaussian decomposition in green. The model is made by the addition of the emission produced by the narrow and broad components (shown in blue), and an additional broad component required to complete the fit (shown by the red dashed line). Middle: Residual profiles of the emission (middle top) and absorption (middle bottom) fits. Bottom: Weighted-mean absorption profile of the stacked GASKAP non-detections. The original stacked spectrum is shown in black. The narrow and broad components' absorption fits are shown by the blue dashed lines, with their total contribution to the absorption profile shown in the solid blue line.}
    \label{fig: Stack Gaussian Decomposition Non-Detections}
\end{figure}

\subsection{Stacked spectrum}
\label{sec: Stacked Spectrum Non-Detections}

For the non-detection stack, we follow the same stacking analysis used in previous sections for the detection sample in Section \ref{subsec: Stacking Subtracted Spectra}. The top and bottom panels in Fig. \ref{fig: Stack Gaussian Decomposition Non-Detections} show the stacked emission and absorption spectra of the non-detection sample, respectively. The high sensitivity of the absorption stack reveals a two-component profile; a deep narrow primary component resembling absorption profiles of the CNM seen in the detection catalogue, and a broad secondary component.

The rms noise of the stacked absorption profile is $\sigma_{\tau} = 2.0 \times 10^{-3}$, calculated by taking the standard deviation of 30 empty channels in the post-shifting velocity range $-50 < v < -20$\,\kms. The noise level has decreased from the original mean of $\sigma_{\tau} = 1.7\times10^{-1}$ for the non-detections by nearly two orders of magnitude. The absorption stack's noise level is also approximately 15 times lower than that of the median noise level ($\sigma_{\tau} = 3.1 \times 10^{-2}$) of the single lines-of-sight classified as detections in the GASKAP-HI absorption survey \citep{Nguyen+2024}.
The stacked absorption profile has an equivalent width (EW) in the velocity range $-50 < v < 50$\,\kms\ of EW $= \int_{-50}^{50} {(1 - e^{-\tau}) \ dv} = 0.44$\kms \citep{Murray+2014}.

\begin{figure}
	\includegraphics[width=\columnwidth]{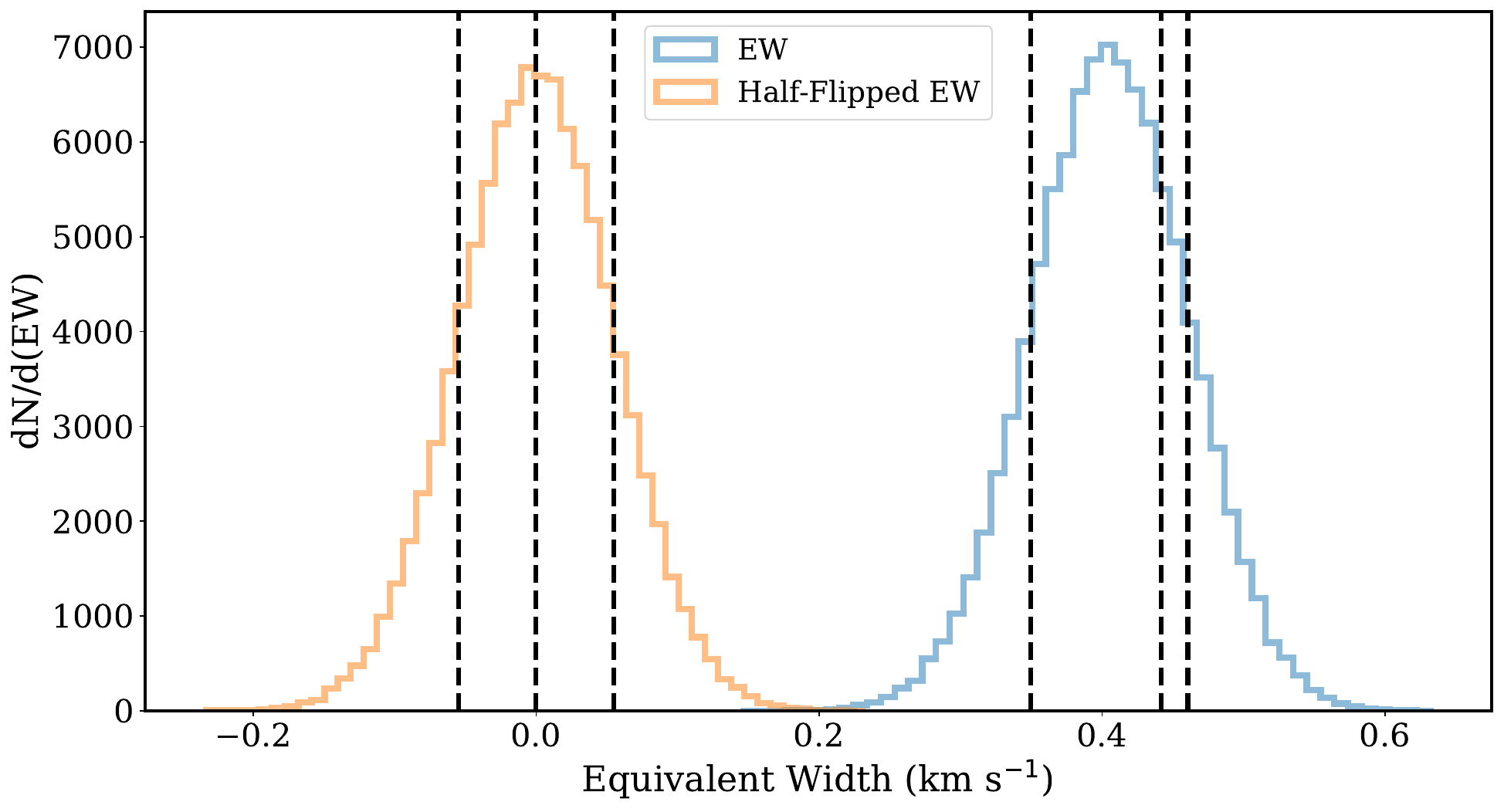}
    \caption{Histograms of the EW distribution from $10^{5}$ trials of bootstrapping, randomly sampling 2240 spectra from the total non-detection sample with possibility of replacement. The blue histogram displays the distribution of EW values from the trials, with a mean of $\langle $EW$ \rangle = \langle \int_{-50}^{50} {(1 - e^{-\tau}) \ dv} \rangle = 0.41 \pm 0.06$\kms\ ($68$ per cent confidence). The orange histogram displays the same random trials except with half of the spectra inverted (multiplied by $-1$) before stacking to test for contamination from outlying spectra. The mean of the orange distribution is $0.00 \pm 0.05$\kms\ ($68$ per cent confidence). The vertical dotted line crossing the middle of the blue histogram represents the EW of the original stack (EW$=0.44$ 
    \kms). The dotted line along the middle of the orange histogram represents the expected mean of $0$ \kms\ for the half-inverted stack EW. The additional dashed lines mark the $68$ per cent confidence interval for the two distributions.}
    \label{fig: EW Distribution}
\end{figure}

To ensure that our stack is not plagued by any outlying spectra and that the primary component of the stack is not produced by only a few dominant narrow profiles in the average, we conducted the same two bootstrapping Monte Carlo simulation experiments used in Section \ref{subsec: Stacking Subtracted Spectra}. We completed the stacking analysis on a new sample of 2240 sources randomly chosen from our original 2240 sightlines with replacement. The process was repeated $10^{5}$ times, with the EW computed in each trial. The distribution of EW values is shown in Fig. \ref{fig: EW Distribution}. The distribution is Gaussian, with a mean value of $\langle EW \rangle = 0.41 \pm 0.06$ \kms\ ($68$ per cent confidence), suggesting that the EW computed from the original sample is consistent with being drawn from a parent population of spectra with a comparable mean and is not influenced by outlying spectra.

We produced an additional EW distribution where in each trial we inverted (multiplied by $-1$) half of our randomly selected sample. The distribution, also shown in Fig. \ref{fig: EW Distribution}, is centred around $0$\kms, consistent with no signal. From this result, we conclude our stack does not produce a false signal.

Since there are multiple components within this stacked absorption spectrum, we followed the method described in Section \ref{sec: Stacked Spectrum Detections} to decompose the spectra and calculate the gas' properties, tabulated in Table \ref{table: Spatial Bins}. The primary and secondary component of the total stacked spectrum using all 2240 sources have a mean spin temperature of $T_s=98\pm12$\,K and $T_s=255\pm 106$\,K, respectively. The uncertainties were evaluated by repeating the same fitting procedure on the $10^{5}$ trials of the bootstrapping sample used above. These uncertainties are larger than those from the variation of the fraction of background warm gas in the radiative transfer solution.

\subsection{Spatial distribution of stacked features}
\label{subsec: Spatial Distribution Stack}

Given the large number of sources that have been obtained within GASKAP's relatively narrow field of view of the Magellanic system, we have the unique opportunity to search for any spatial variation in the absorption stack profile. In this work, we have chosen to bin the data by increasing source density of detections, allowing us to assess how the profile's peak optical depth and spin temperature change with distance from known locations of CNM gas. Since regions of cold gas typically show higher column densities, this also acts as a probe for binning by column density (as shown in Table \ref{table: Spatial Bins}).

\begin{table*}
\begin{center}
\centering
\begin{tabular}{l rrrrrrr}
\hline\hline
Bin &  N$^{a}$ & n$^{b}$      & N$_{\rm tot}$                 & $\sigma_{\tau}$ $^{c}$  &  Peak $\tau$        & FWHM   & $T_s$ \\
    &    & (deg$^{-2}$) & ($\times10^{20}$ cm$^{-2}$)   &  ($\times10^{-3}$) &  ($\times10^{-3}$)  & (\kms) & (K)  \\
\hline
Total & 2240   & 0.97$\pm$0.76  & 4.2 & 2.0 & 60$\pm$10 &  4.0$\pm$1.1  & 98$\pm$12  \\
Non-detections &     &    &        &       & >20$\pm$10** &  <14.2$\pm$1.9** & 255$\pm$106 \\
\hline
1 & 374 & 0.08$\pm$0.06 & 2.5 & 7.0 & - * & - *  & - *  \\
  &     &               &      &     &     &      &    \\
2 & 373 & 0.32$\pm$0.09 & 3.1 & 7.8 & 46$\pm$20  & 3.0$\pm$1.5  & 75$\pm$26  \\
  &     &               &      &     & >17$\pm$6**  & <13.5$\pm$3.6** & 306$\pm$117 \\
3 & 373 & 0.66$\pm$0.11 & 4.0 & 7.8 & 87$\pm$9  & 4.1$\pm$1.8  & 77$\pm$26  \\
  &     &               &      &     &     &      &    \\
4 & 373 & 1.02$\pm$0.12 & 4.8 & 7.8 & 62$\pm$27  & 2.7$\pm$1.5  & 89$\pm$24 \\
  &     &               &      &     & >40$\pm$15**  & <10.2$\pm$4.3** & 135$\pm$215 \\
5 & 373 & 1.48$\pm$0.15 & 5.5 & 8.1 & 107$\pm$10 & 5.6$\pm$0.7  & 99$\pm$7  \\
  &     &               &      &     &     &      &     \\
6 & 374 & 2.28$\pm$0.34 & 6.2 & 12.2 & 187$\pm$16 & 6.1$\pm$0.3  & 88$\pm$6  \\
  &     &               &      &     &     &      &    \\
\hline
\end{tabular}
\captionsetup{justification=centering}
\caption{Fitted parameters for the total non-detection stack and spatially binned stacks \\
\textbf{Notes:} \\
$^{a}$Number of non-detection sources. \\
$^{b}$Mean plane-of-sky number density of detection sources. \\
$^{c}$Mean rms noise over 30 velocity channels from approximately $-50$ to $-20$ \kms. \\
$^{*}$Due to the low signal-to-noise of the data in bin 1, a fit could not be made. \\
$^{**}$As discussed in Section \ref{sec: Stacking Considerations}, optical depths and FWHM of the secondary components in shifted stacks are lower limits and upper limits respectively.}
\label{table: Spatial Bins}
\end{center}
\end{table*}

\subsubsection{Binning by source density of detections}

\begin{figure}
	\includegraphics[width=\columnwidth]{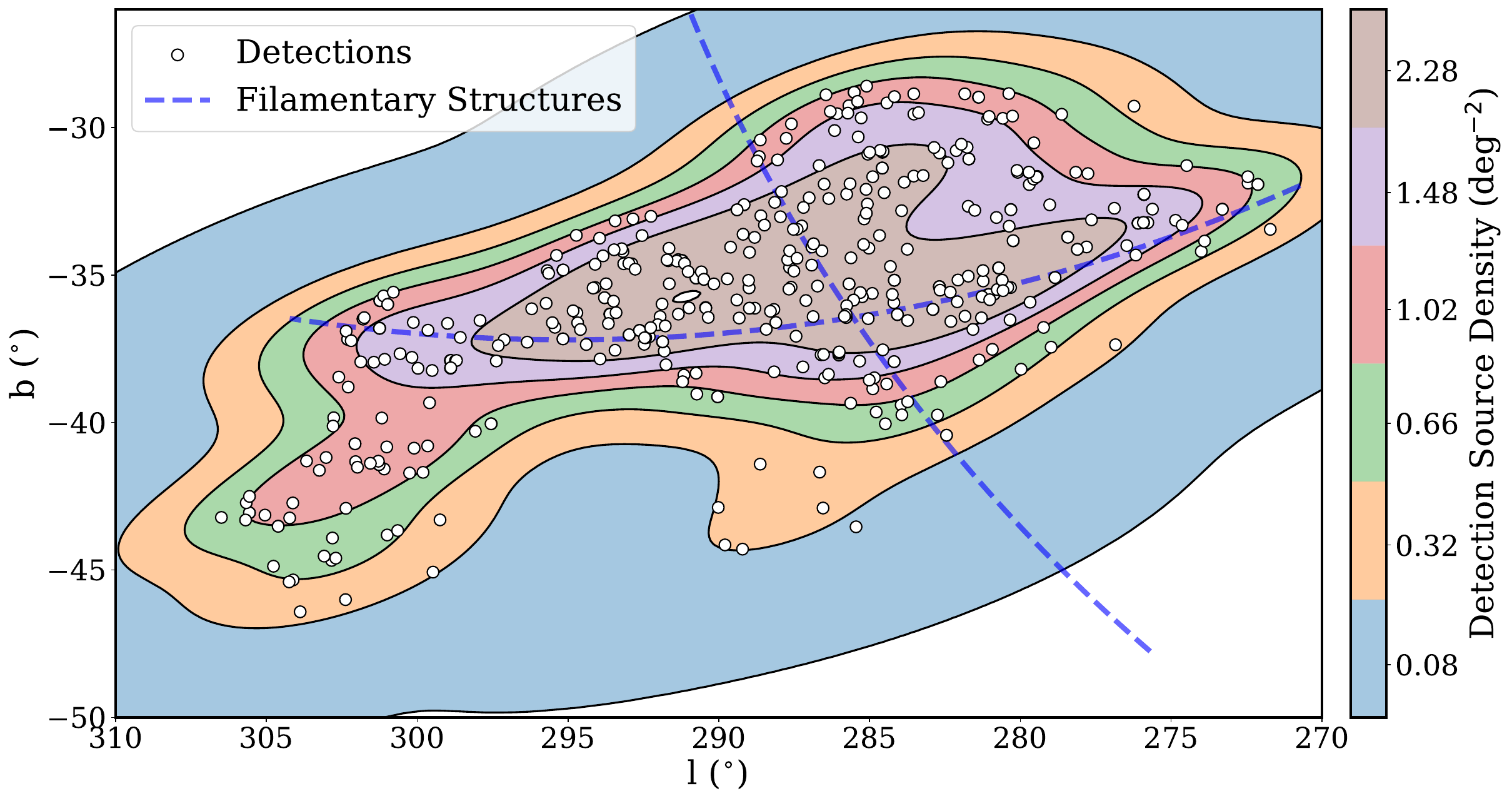}
    \caption{A 2-dimensional contour plot of the source density of Milky Way \HI\ detections found through the GASKAP Pilot II survey towards the Magellanic system \citep{Nguyen+2024}. The contours are created based on a kernel density PDF using the spatial location of the detected sources (shown with white circles). The levels of the contours separate the 6 equally sized bins of non-detection sources, highlighting regions of higher absorption source density on the plane-of-sky towards the centre of the \HI\ filamentary structure observed in Fig. \ref{fig: Field of View of Non-Detection Sources}. The dotted blue lines indicate the \HI filaments from Fig. \ref{fig: Field of View of Non-Detection Sources} for context.}
    \label{fig: Density of Cold Components}
\end{figure}

\begin{figure}
	\includegraphics[width=\linewidth]{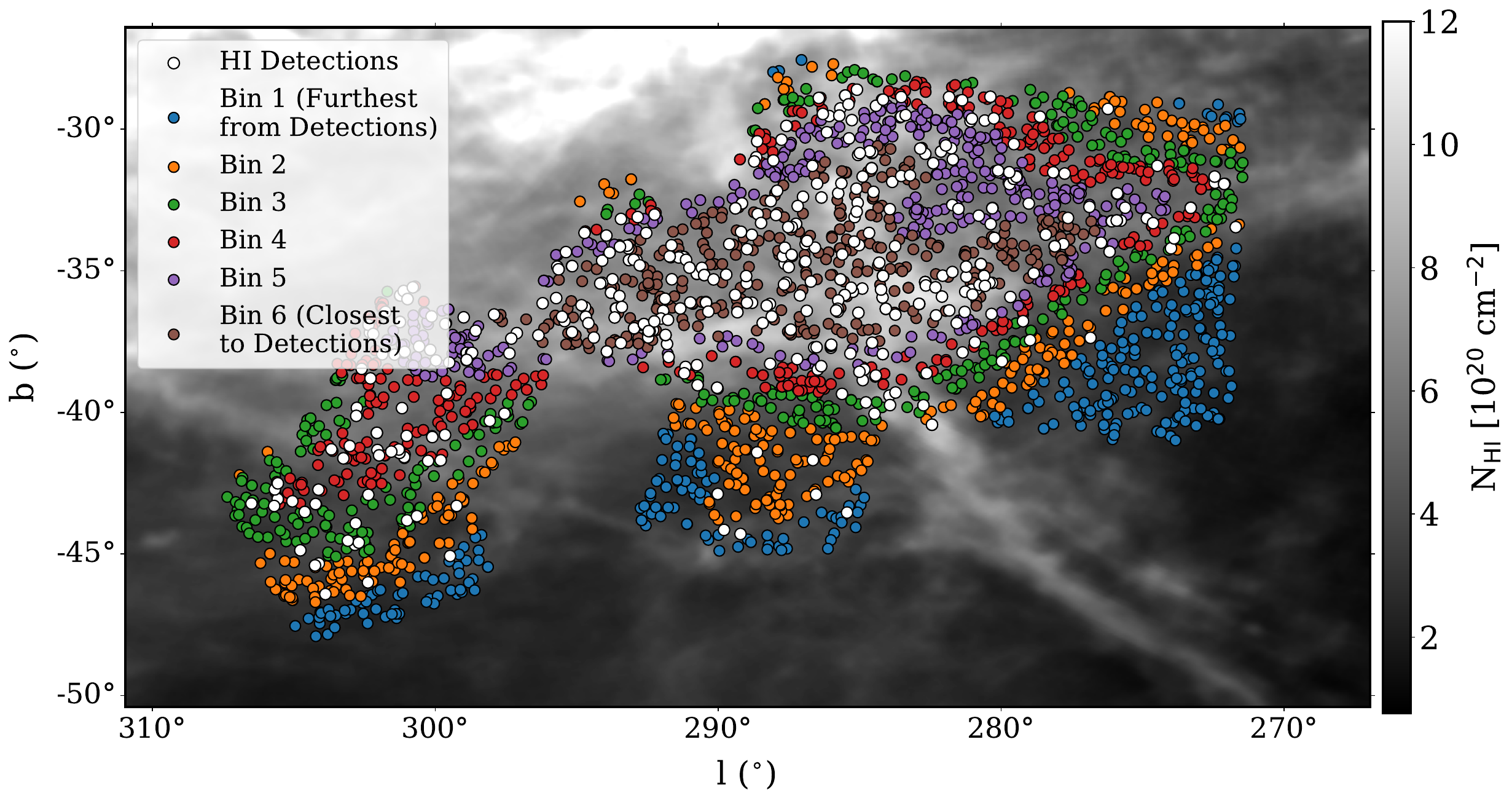}
    \caption{Spatial distribution of the non-detection sources coloured based on their respective contour level from the contour regions shown in Figure \ref{fig: Density of Cold Components}. We define the bins $1-6$ starting from the outer regions of the survey, with higher bin numbers probing regions with higher source density of \HI\ absorption, as highlighted with the locations of the 462 Milky Way \HI\ absorption detections marked with white circles. The background is a grey-scale of the Milky Way \HI\ column density in the region of the sky from the GASS survey \citep{Mcclure-Griffiths+2009,Kalberla+2010,Kalberla_Haud+2015}.}
    \label{fig: Spatial Bin of Non-Detections}
\end{figure}

One parameter of interest is how the profile's peak optical depth and spin temperature change with distance from known locations of CNM gas. To probe this, we first used the locations of the 462 \HI\ detections of Milky Way absorption in this region \citep{Nguyen+2024} to create a kernel probability distribution function (PDF) of known cold gas. We convert the density values of the PDF into source density of detections in units of sources per square-degree. We attribute to each non-detection a value corresponding to the source density grid at the non-detection's spatial location. 

The non-detections were then binned into six equally sized populations based on their respective PDF values. The six regions created from this binning process are displayed via the contour map in Fig. \ref{fig: Density of Cold Components}, overlaid by the location of absorption detections shown in white used to calculate the source density map and the filamentary structures marked in Figure \ref{fig: Field of View of Non-Detection Sources} for context.

The spatial distribution of the individual non-detection sources after binning is displayed in Fig. \ref{fig: Spatial Bin of Non-Detections}, overlaid onto the column density map from the GASS \HI\ emission survey. Binning using this source density metric highlights a concentric diamond-configuration centred on the Galactic filament cross, marking it as a region densely populated by CNM detections. With each of these bins of non-detections, we repeated the same methodology of stacking their absorption and emission spectra and fitting them using Gaussian decomposition as done in Sections \ref{subsec: Stacking Subtracted Spectra} and \ref{sec: Stacked Spectrum Non-Detections} to extract their physical properties. We repeated this using the same Monte Carlo bootstrapping process as done previously to determine uncertainties. 

\begin{figure*}
	\includegraphics[width=0.32\linewidth]{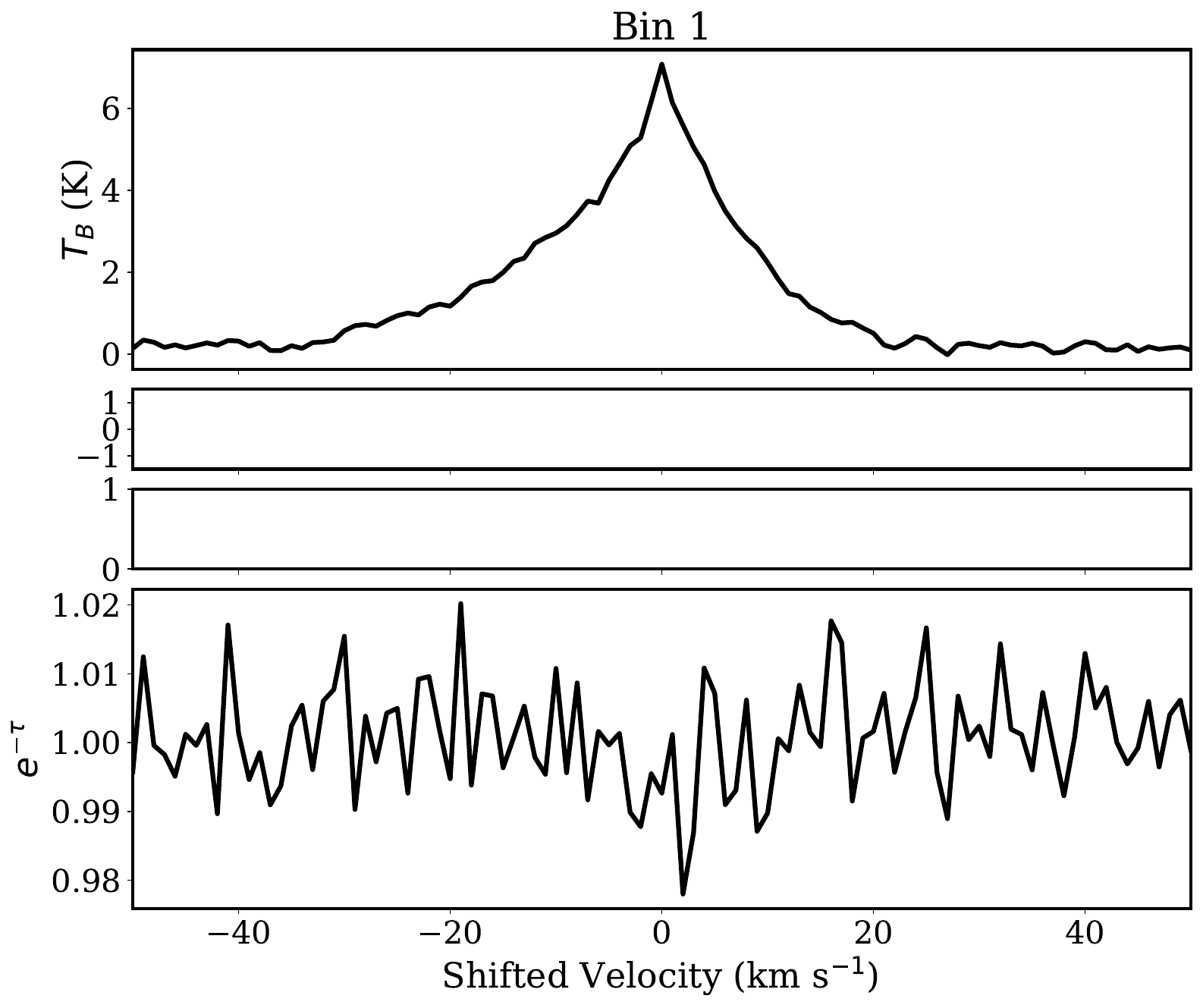}
	\includegraphics[width=0.32\linewidth]{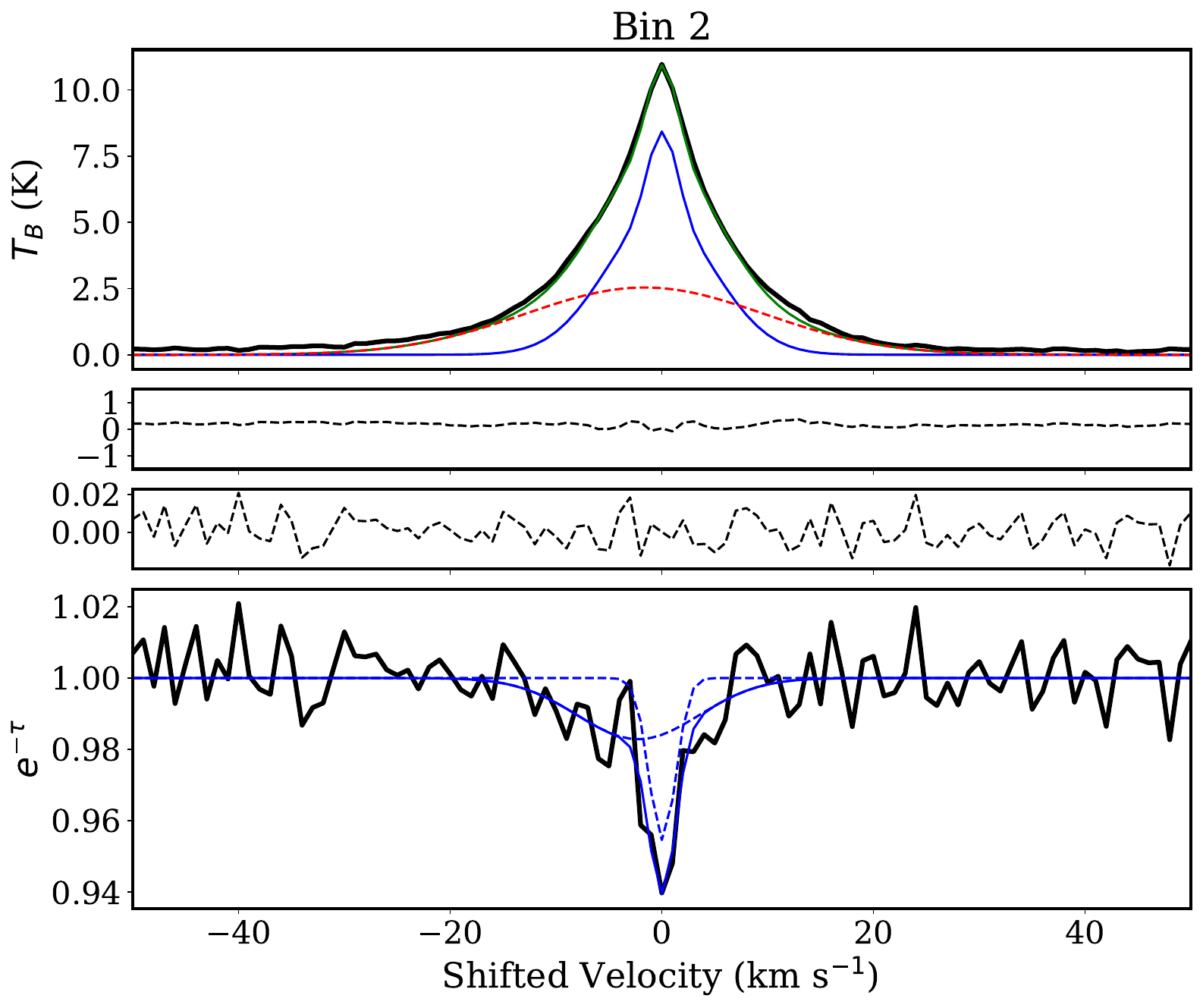}
	\includegraphics[width=0.32\linewidth]{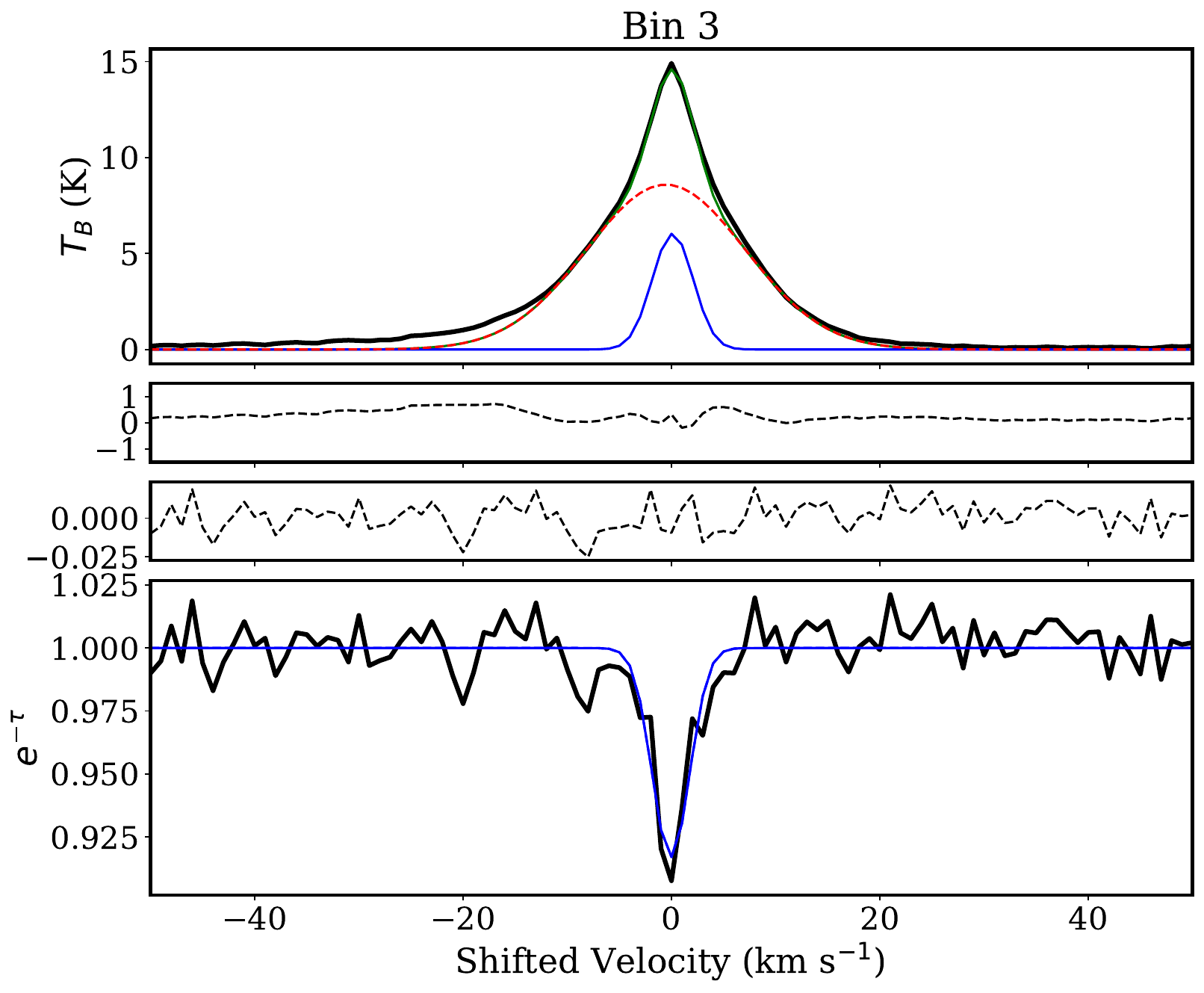}
	\includegraphics[width=0.32\linewidth]{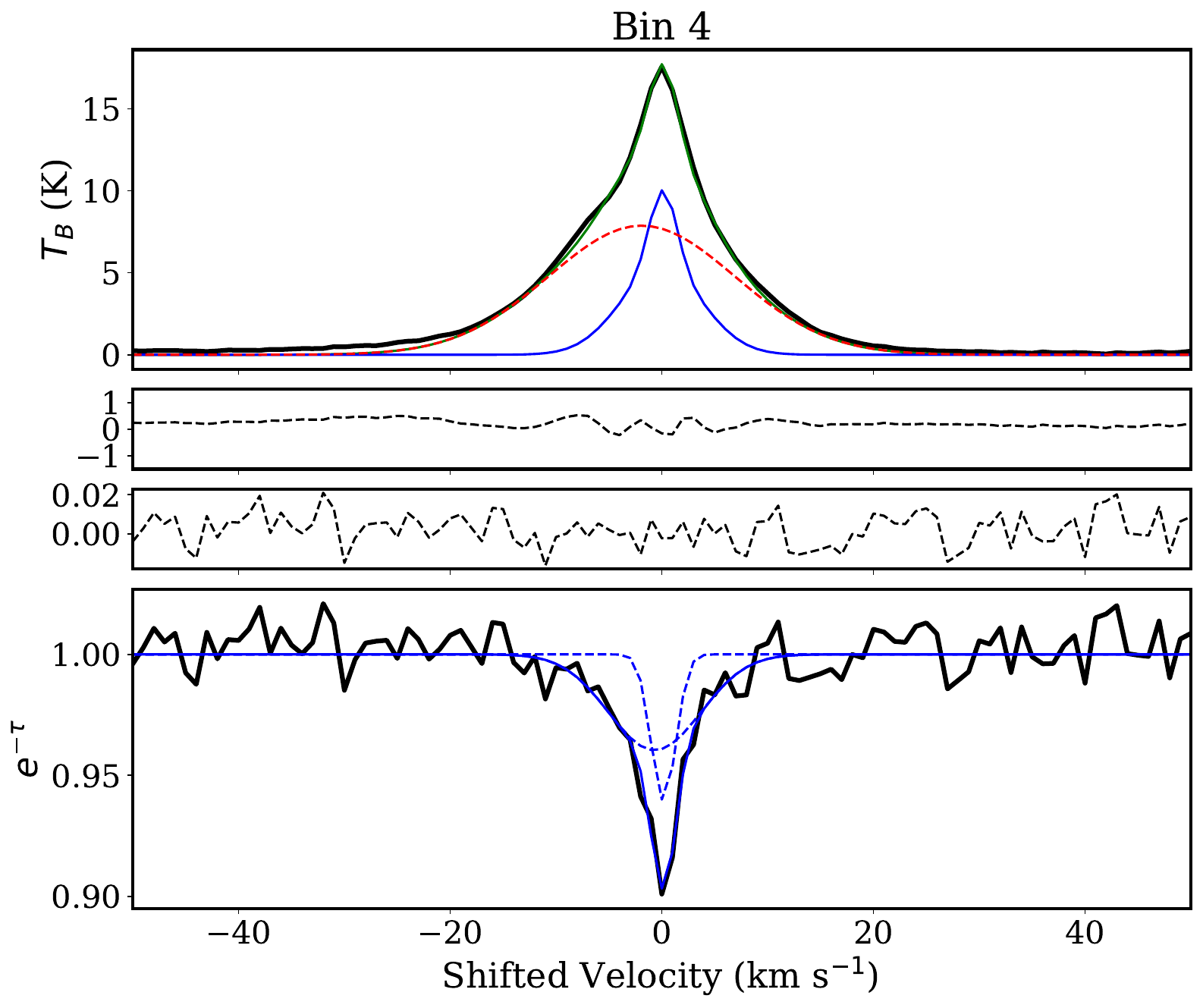}
	\includegraphics[width=0.32\linewidth]{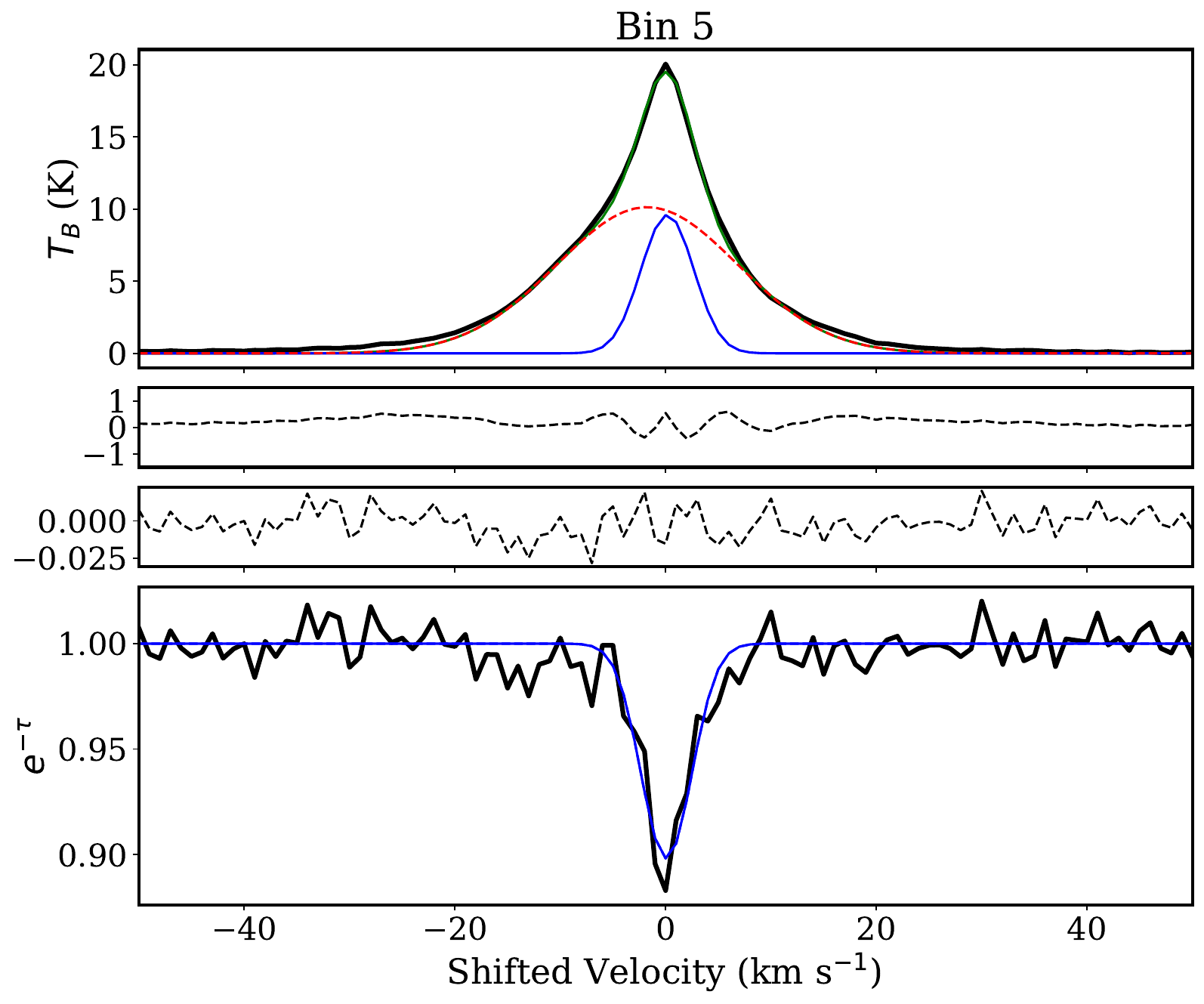}
	\includegraphics[width=0.32\linewidth]{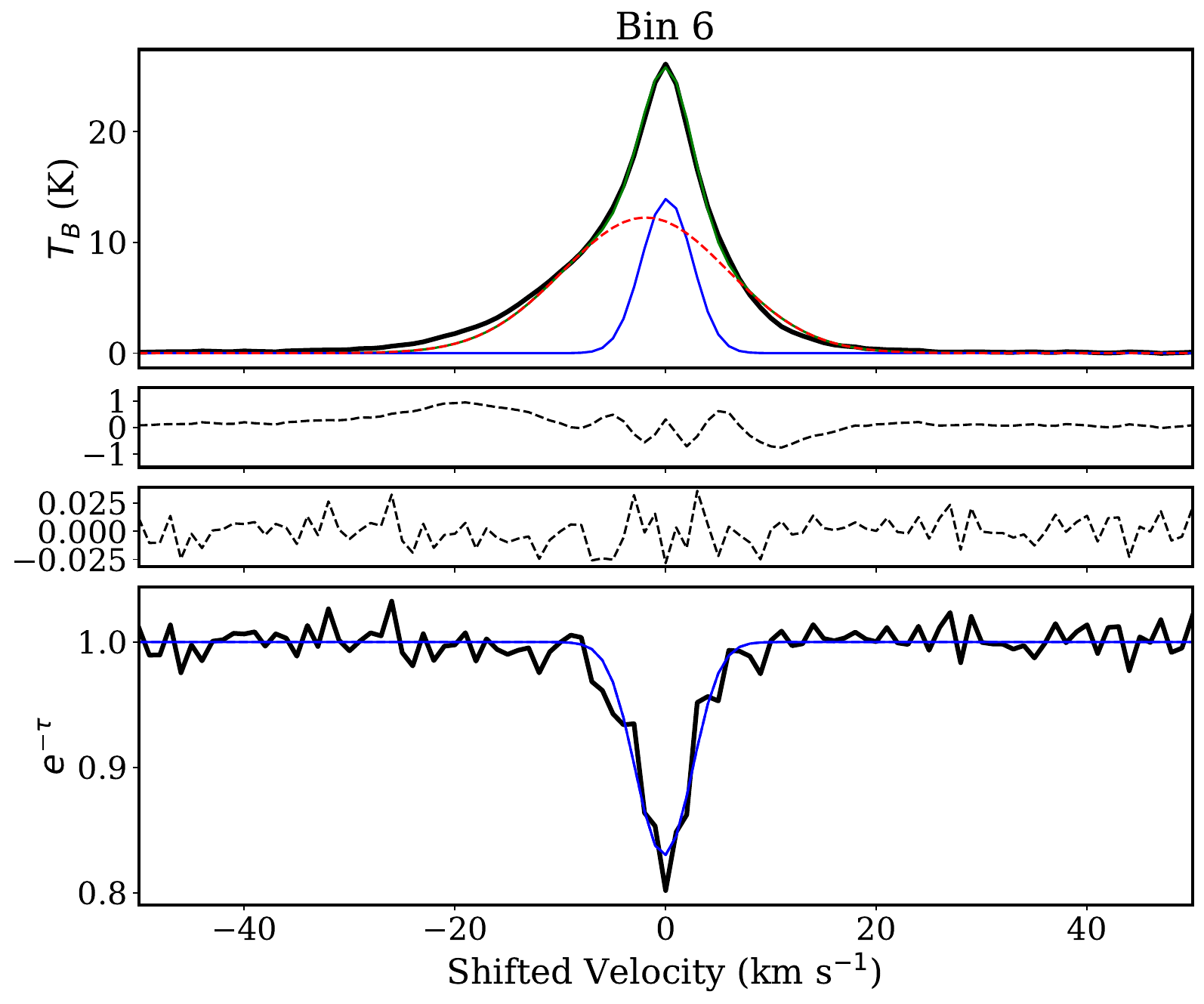}
    \caption{\HI\ emission and absorption stacked profiles of the spatially binned groups of spectra from Fig. \ref{fig: Spatial Bin of Non-Detections} and their Gaussian decompositions using the method from \citet{Heiles_Troland+2003}. The layout and colour scheme used is similar to that used in Fig. \ref{fig: Stack Gaussian Decomposition Non-Detections}. We note a secondary component is detected in bins 2 and 4, whereas the primary component exists in all bins except for bin 1, increasing in amplitude in absorption from bins 2 to 6 with increased detection source density. Due to the signal-to-noise being inconsistent between bins, bin 1 could not be fitted with a statistically significant Gaussian component.}
    \label{fig: Gaussian Decompositions of Spatial Bins}
\end{figure*}

The Gaussian decomposition of each bin is shown in Fig. \ref{fig: Gaussian Decompositions of Spatial Bins}. Table \ref{table: Spatial Bins} displays the physical properties obtained including the total column density of the bins, $N_{\rm tot} \approx 1.823 \times 10^{18} \int{T_{\rm B} \ dv}$, peak optical depth, spin temperature and FWHM of each component. We note that with each bin getting closer to regions of cold gas (bin 1 to bin 6 increasing in source density), the total column density also increases, allowing us to also probe the variation of the absorption features with column density simultaneously.

\begin{figure*}
    \begin{multicols}{2}
    \includegraphics[width=\columnwidth]{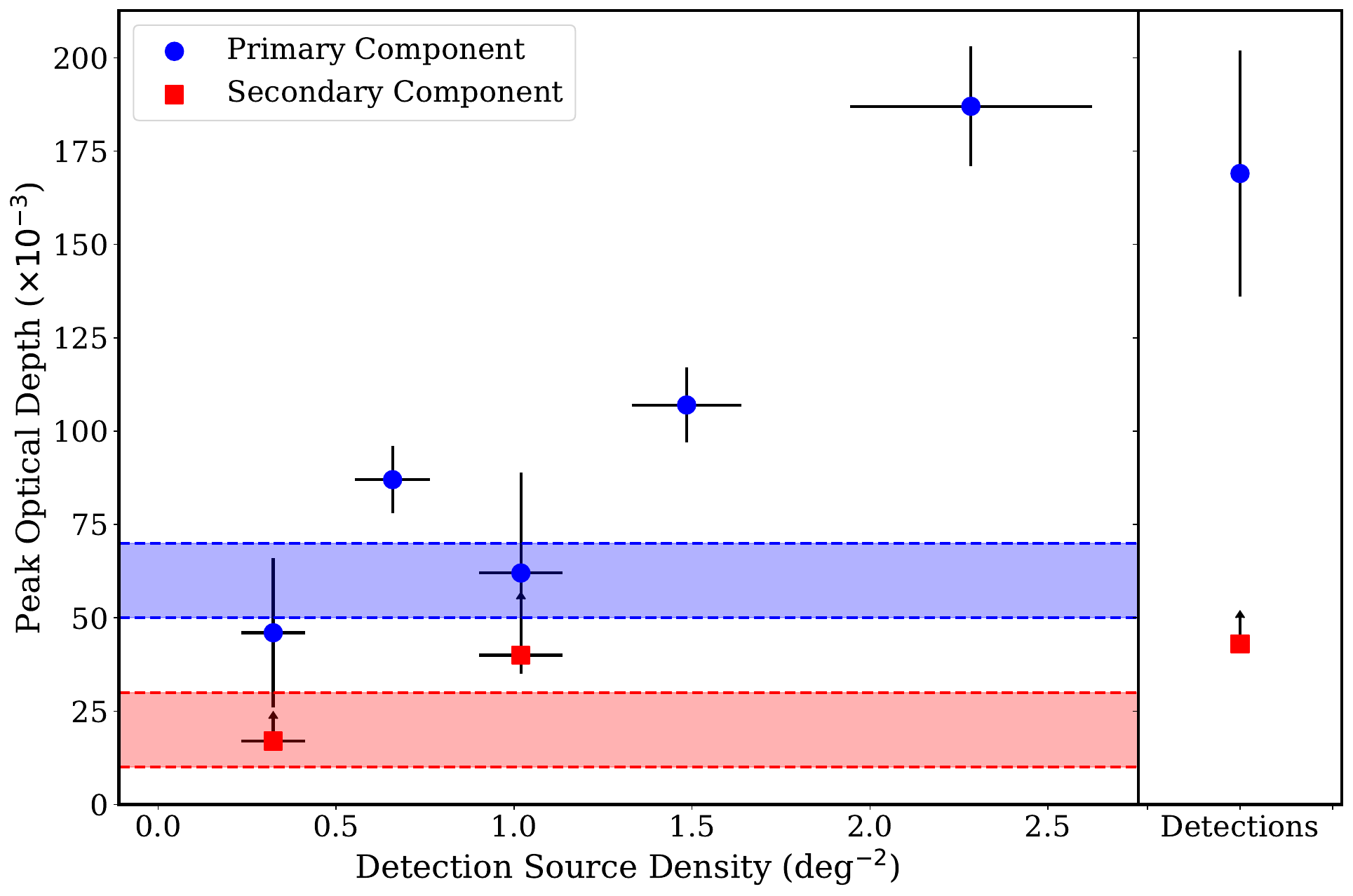}\par
    \includegraphics[width=\columnwidth]{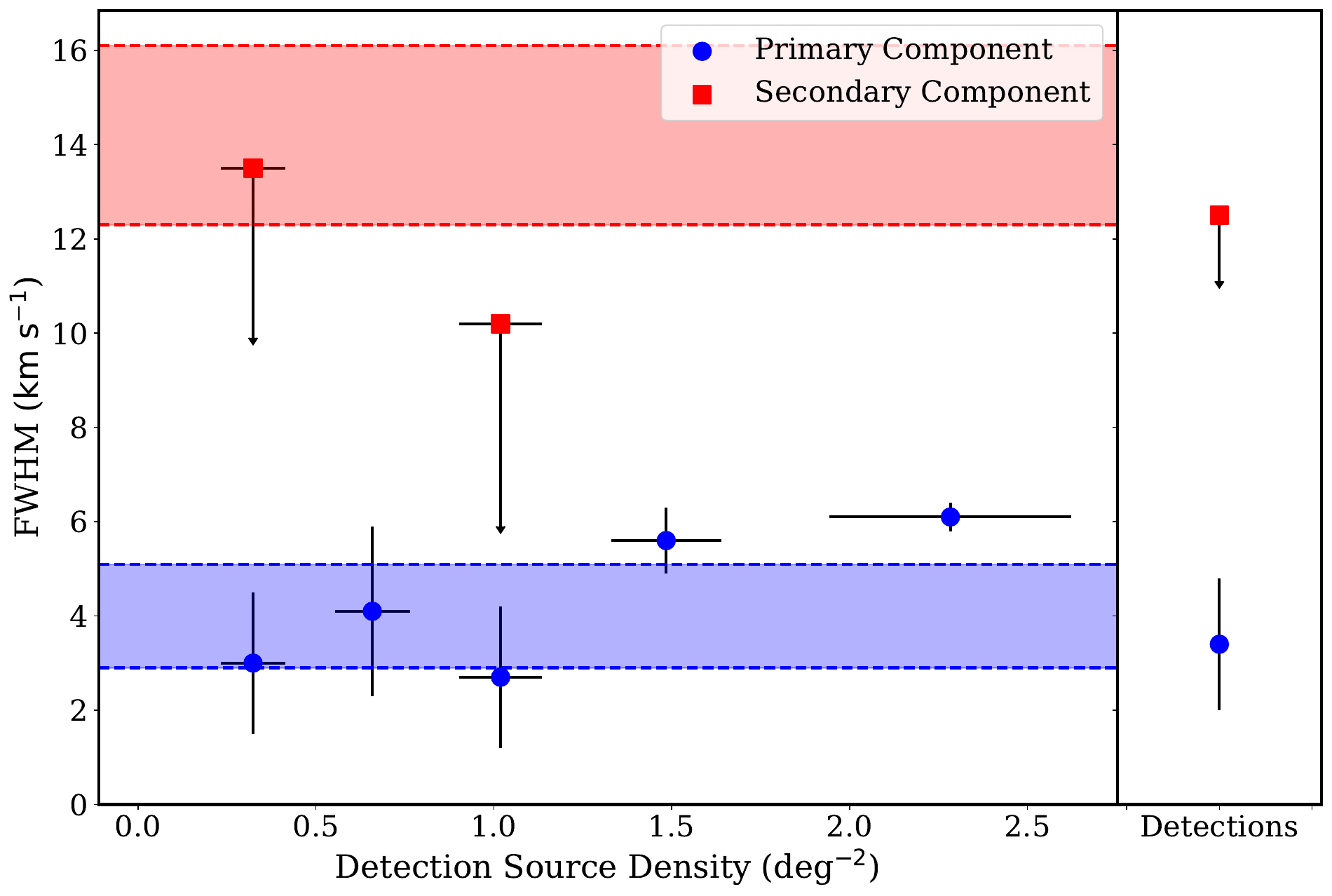}\par 
    \end{multicols}
    \caption{Properties of the Gaussian decompositions of the stacked profiles binned in Section \ref{subsec: Spatial Distribution Stack} (bin 1 not shown). Left: Plot of the peak $\tau$ values for the primary components (blue circles) and secondary (red squares) components from each of the stacked absorption profiles. The primary component increases in peak $\tau$ as the stacked spectra get closer to regions of known cold gas, and increase at a faster rate relative to the secondary component, whose lower limit remains relatively stable in peak optical depth. The blue and red shaded regions represent the peak optical depth of each component in the total stack from Fig. \ref{fig: Stack Gaussian Decomposition Non-Detections} within $1\sigma$. The optical depths of the detection sample stack from Section \ref{sec: Stacked Spectrum Detections} are shown to the right, and agree with the trends from the non-detection bins as they approach colder gas. Right: Plot of the FWHM for the primary (blue circles) and secondary (red squares) components from each of the bins' stacked profiles. The FWHM of the primary component in each bin is consistent with that of the total stack, shown in the shaded region. The secondary component's upper limit of the FWHM is also relatively consistent, except for the outlier in bin 4. The FWHM's of the detection sample stack from Section \ref{sec: Stacked Spectrum Detections} are shown to the right, and agree with values of those from the non-detection bins.}
    \label{fig: Properties In Each Spatial Bin}
\end{figure*}

The primary component's peak optical depth diminishes from approximately $\tau_{\rm peak} = (187 \pm 16) \times 10^{-3}$ to $(46 \pm 20) \times 10^{-3}$ as the bins of non-detections move outwards, away from regions of known cold gas, before disappearing in the bin most distant from the GASKAP detections. This trend is supported further with the left plot of Fig. \ref{fig: Properties In Each Spatial Bin} highlighting the variation of peak $\tau$ of the primary component in each bin, as it decreases the further we move from the filament structure. The peak optical depth values from the detection sample stack in Section \ref{sec: Stacked Spectrum Detections} is also displayed, representing the closest available bin in our GASKAP sample to cold gas. We find the detection stack value agrees with the trend of the primary components in the non-detection bins.

The secondary component values on the other hand appear to plateau in peak optical depth, with the value in bin 4 being relatively similar to that of the detection sample stack in the central region of the filament structure. While this is a lower limit of the true average (as discussed in Section \ref{sec: Stacking Considerations}), the plateau is still noteworthy considering these lower limits are linearly correlated with their true counterparts (as shown in Appendix \ref{sec: Toy Model Theoretical Considerations}).

We note that due to variation of signal-to-noise between each bin, bin 1's signal-to-noise was too low to consistently Gaussian decompose the stack during the Monte Carlo process used in this work. As a result, uncertainties could not be calculated for this bin. Therefore, we treat bin 1 as a non-detection of \HI absorption.

The right panel of Fig. \ref{fig: Properties In Each Spatial Bin} shows the FWHM of the two absorption components in each bin. We note the primary component's FWHM is consistent with that of the total stack. The secondary component's FWHM, while an upper limit, also seems consistent with the value of the total stack (except for an outlier in bin 4). With the flux of these stacked components being conserved, the FWHM upper limit is related to the lower limit of the peak optical depth. As such, this implies that the FWHM upper limit is also linearly correlated with the true average FWHM (similar to the optical depth as mentioned earlier). Therefore, while we can not confirm that the true FWHM in each bin is similar to the total stack, we can conclude that the bins' values are at least similar to each other.

\begin{figure}
    \includegraphics[width=\linewidth]{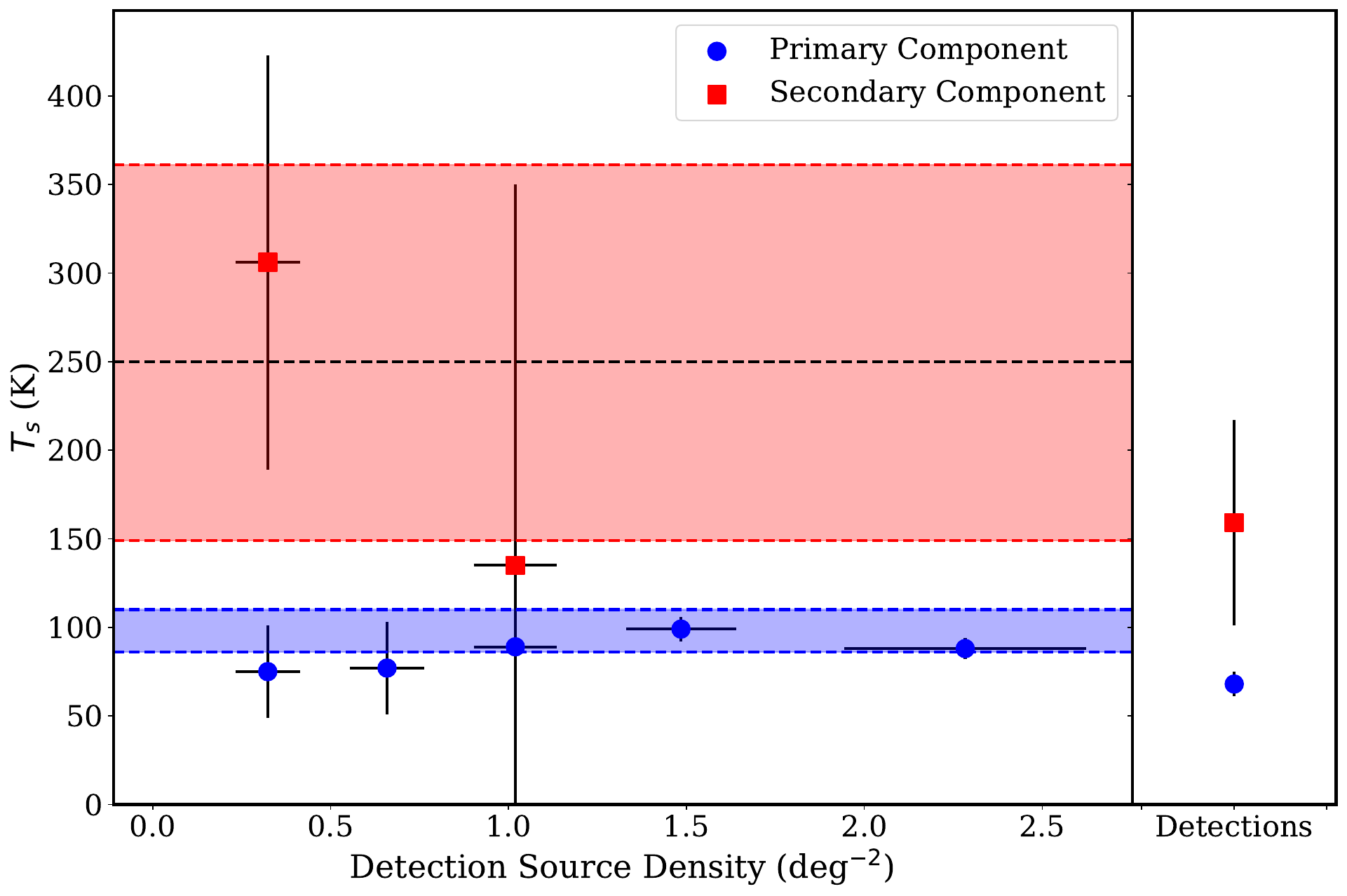}\par 
    \caption{Variation of the spin temperature values for the primary (blue circles) and secondary (red squares) components from each of the stacked profiles binned in Section \ref{subsec: Spatial Distribution Stack} (bin 1 not shown). The black line denotes the cutoff between CNM and UNM used in this work of $\sim250$ K. The primary component remains relatively consistent  in spin temperature regardless of the region being probed. Both components align with their counterparts in the total stack, shown in the shaded regions with their widths indicating the $1\sigma$ standard deviation. The spin temperatures of the detection sample stack from Section \ref{sec: Stacked Spectrum Detections} are shown to the right, and align with those from the non-detection bins.}
    \label{fig: Spatial Bins Tspin}
\end{figure}

Fig. \ref{fig: Spatial Bins Tspin} displays the variation of $T_s$ for the primary and secondary components with distance from cold gas. We find that both the primary and secondary components remain relatively consistent in temperature regardless of distance or changes in detection source density, and notice no overall trend. There is however a minor decline in the secondary component's spin temperature as they approach the value of the detection stack from Section \ref{sec: Stacked Spectrum Detections}. Regardless, the spin temperatures of the two components still align well with those of the total stack, with most bins' components falling within $1\sigma$ (68 per cent confidence) of the total stack's values. One might argue that the stability in spin temperature could be caused by the additional component added in emission during the fitting process accommodating any changes in the absorption properties. However, since the linewidths of the broad and narrow components (see values tabulated in Table~\ref{table: Spatial Bins}), and hence their $T_{k,\rm max} = 121 \sigma_{v}^{2}$ remain relatively consistent, similar to their spin temperatures, it is likely that this trend of constant spin temperature is physical and not a byproduct of the fitting process.

\section{Discussion}
\label{sec: Discussion}

Previous absorption surveys have typically used deep observations in single LOS locations \citep{Carilli+1998,Dwarakanath+2002,Patra+2018} or observations in a widespread field \citep{Heiles_Troland+2003,Murray+2018}. In this work however, we have taken advantage of the unbiased observational capabilities of ASKAP to observe the spatial distribution of the properties of cirrus \HI\ gas in the relatively small region of the Milky Way toward the Magellanic system.

\subsection{Spin temperature}

In Section \ref{sec: Stacking Model-Subtracted Detections}, stacking the residual detection spectra reveals a broad component with a spin temperature of $1320\pm263$ K. This is representative of gas from the UNM, and is much higher than the highest temperature detected by \cite{Nguyen+2024} from the 462 individual detection sightlines. Since this is a stacked average of the detection spectra, which are sampled in an unbiased manner across the spatial span of the GASKAP survey range, this component represents an average of the UNM temperature within this region of the sky. The component's spin temperature is consistent with average UNM temperatures found from previous surveys that observed relatively diffuse gas ($\sim800\pm570$ K from \citealp{Heiles_Troland+2003} and $\sim600\pm480$ K from \citealp{Murray+2018}).

The fact that this broader UNM component appears simultaneously in the same region of sky as the cold components found in both \cite{Nguyen+2024} as well as this work supports previous findings that CNM structures are typically surrounded by a broader warmer feature such as colder UNM \citep{Kalberla+2018} and that they are interlinked as part of a continuous range of \HI\ properties, with the UNM being a transitional stage between the stable phases. 

From the spin temperatures of the components in the total non-detection \HI\ stack in Fig. \ref{fig: Stack Gaussian Decomposition Non-Detections}, we are able to discern that both the primary and secondary components with temperatures of $98\pm 12$ and $255\pm 106$ K originate from gas belonging to the CNM ($T_s < \sim250$ K). We note that both values are higher than the mean spin temperature of $\sim 50$\,K for the components found in single lines-of-sight \citep{Nguyen+2024}. This discrepancy is likely due to the sightlines used being non-detections and as such did not contain deep absorption profiles of colder denser gas like those classified as detections. However, only a few instances of warmer gas still within the regime of CNM and under $250$\,K were found in the detection survey. This upper regime of cold gas has optical depths beyond the sensitivity of the current Pilot survey and as such were only found through stacking in this work to reduce the signal-to-noise. The primary component's spin temperature also aligns with the average CNM spin temperature of $80\pm60$ K from the Millenium survey \citep{Heiles_Troland+2003} and $71\pm52$ K from the more recent 21-SPONGE survey \citep{Murray+2018} albeit slightly warmer. This indicates that while probing more diffuse gas than typical past \HI absorption surveys, the CNM spin temperatures remain relatively consistent.

The secondary component's higher spin temperature of a warmer CNM gas seems to indicate that the sensitivity reached by the stacked spectrum is high enough to detect and quantify the properties of warmer cool gas throughout the entire region of the GASKAP survey. As shown in Section \ref{subsec: Multiple Components}, while the observed spin temperature of primary and secondary components is well correlated with the true average, we do not have any information as to the dispersion of the individual spin temperatures that form these averages.
Due to the range of thermal pressures allowing a two-phase medium, the CNM temperature predicted can theoretically range from $\sim20$ to $350$ K in Solar neighbourhood conditions \citep{Wolfire+2003}. While the upper limit of stable spin temperatures for the CNM is a function of dust extinction \citep{McClure-Griffiths+2023}, we can only safely conclude that this secondary component at approximately $250$ K is at the rough boundary of the cold and unstable regime of the neutral ISM. We conclude that the secondary component likely contains a mixture of warmer CNM and cooler UNM.

Unlike the detection sample stack, the non-detections when averaged did not reveal any gas with temperatures above $\sim1000$ K. This is likely because the UNM component's optical depth found in the detection sample residual stack resides close to the $\sim3\sigma$ detection threshold with the non-detection stack's noise level. Since the detection and non-detection sightlines are close in location and span the same small patch-of-sky, it is likely such a component could be found with the non-detections with longer exposure time and hence a reduced noise profile.

The WNM has been observed in absorption in the past \citep{Carilli+1998,Murray+2014,Patra+2024}; however, our large sample of absorption signals when stacked only yield components with spin temperatures belonging to the CNM in the non-detection stack and UNM in the detection stack. This is in contrast to recent work observing WNM temperatures up to $~10^{4}$ K in \HI absorption at optical depth sensitivities close to that of the absorption stack in this work at $\sigma_{\tau} \approx 1 - 2 \times 10^{-3}$ \citep{Patra+2024}. This could point to a lack of WNM in this region relative to other regions on the sky, especially with the field dominated by 2 large cold dense filamentary structures (\citealp{Nguyen+2024}; Lynn et al., in preparation).
Another possibility is that the WNM in this region resides at higher optical depths than normal. Adding to this is the optical depth sensitivity of our absorption profile stack being an order of magnitude lower than that of more sensitive observations like the 21-SPONGE stacking analysis of \cite{Murray+2018} ($\sigma_{\tau}\approx10^{-3}$ versus the 21-SPONGE stack sensitivity of $\sigma_{\tau}\approx2\times10^{-4}$).
Despite not detecting the WNM in absorption, we can define an upper limit on the peak optical depth of the WNM in this region of the sky of approximately $\tau_{\rm peak, WNM} \lesssim 3.3\times10^{-3}$ using the $\sim 3\sigma$ detection threshold of the fitting algorithm when applied to the residual detection spectra stack in Section \ref{sec: Stacking Model-Subtracted Detections}.

\subsection{Non-detections spatial variations}

To determine whether the temperature and optical depth trends observed in Section \ref{subsec: Spatial Distribution Stack} reflect changes in the true density and relative abundance of the \HI\ phases, we looked to dust. Since at high Galactic latitudes the total amount of hydrogen is in the neutral phase, dust is a good tracer of \HI\ at high latitudes \citep{Low+1984,Boulanger+1996,Lenz+2017}. Converting dust extinction densities from three-dimensional (3D) dust maps to total hydrogen densities (dominated by \HI) therefore allows us to probe any variations to 3D \HI\ density between each of the 6 bins defined in Figs \ref{fig: Density of Cold Components} and \ref{fig: Spatial Bin of Non-Detections}.

Recently, work by \cite{Edenhofer+2024} has produced a local 3D dust extinction map ranging from $69$ pc to $1.2$ kpc from the Sun with an angular resolution of $14'$ and a varying distance resolution, from $0.4$ to $7$ pc. 
We used the coordinates of the sightlines used in the non-detection absorption stacks and collected, using the {\fontfamily{qcr}\selectfont dustmaps}\footnote{\url{https://dustmaps.readthedocs.io/}} package, all of their respective sightlines in the 3D dust map for each of the six spatial bins used in Section \ref{subsec: Spatial Distribution Stack}. 

\begin{figure}
    \includegraphics[width=\columnwidth]{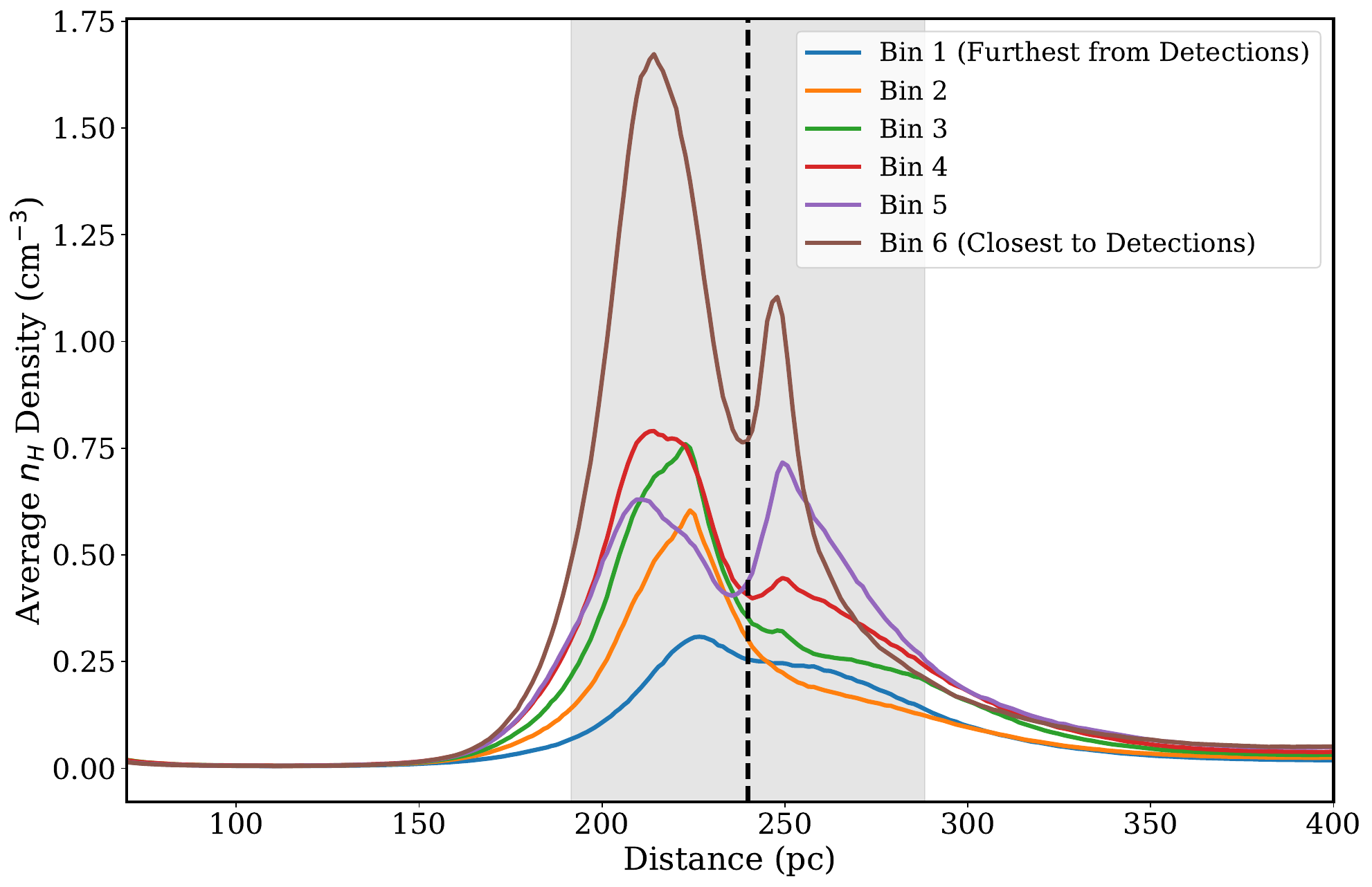} 
    \caption{Weighted-mean of $n_{\rm H}$ along the sightlines of each of the 6 bins described in Section \ref{subsec: Spatial Distribution Stack}. The dust densities were converted from the \citet{Edenhofer+2024} 3D dust map to total hydrogen densities using a conversion factor of $1653$\,cm$^{-3}$ \citep{ONeill+2024}. Note as the spatial bins sample sightlines further away from the known regions of cold gas (from bin 6 to 1), we observe a decrease in the peak density of total hydrogen and by extension a decrease in the amount of colder phases of \HI\ as we are observing at high Galactic latitudes \citep{Lenz+2017}. The vertical dashed line represents the average distance of maximum $n_{\rm H}$ in the 2240 non-detection sightlines ($\sim240 \pm 45$ pc), with the grey region indicating the $1\sigma$ standard deviation.}
    \label{fig: Edenhofer Average Dust Spectra}
\end{figure}

We followed the procedure described in \cite{ONeill+2024} to convert the differential extinction measurements $A'_{ZGR_{23}}$, defined in term of unitless extinction $ZGR_{23}$ \citep{Zhang+2023}, into densities of total hydrogen atoms $\rm H$ (molecular gas included) given by
\begin{equation}
\label{equation: Dust to Hydrogen Conversion}
    n_{\rm H} = C \times A'_{ZGR_{23}} \, ,
\end{equation} 

where $C=1653$\,cm$^{-3}$. The binned $n_{\rm H}$ sightlines were then averaged, weighted by the same $1/\sigma_{\tau}$ values as used in the stacks of Section \ref{subsec: Spatial Distribution Stack}. These weighted-mean $n_{\rm H}$ spectra are shown in Fig. \ref{fig: Edenhofer Average Dust Spectra}. 

We find that the collections of sightlines further away from regions of known cold gas and with a lower detection source density (i.e., bin 1) contain a noticeably lower peak with lower $n_{\rm H}$ densities compared to the regions close to the filament and known cold gas (i.e., bin 6) with the density spectra's peaks increasing in amplitude the higher the detection source density. 
This trend supports our findings in Section \ref{subsec: Spatial Distribution Stack} where the cold gas optical depth disappears drastically as we move away from the filaments (shown in the left plot of Fig. \ref{fig: Properties In Each Spatial Bin}). It is interesting to note that the $n_{\rm H}$ spectra contain two local maxima, implying the existence of multiple physically separated layers of \HI, especially in the bins closer to the location of cold gas. This result aligns well with the fact that absorption sightlines in the region contain multiple cold clouds along individual sightlines and that the region contains two distinct filament structures at different physical distances \citep{Nguyen+2024}.

The fact that the optical depth of the primary cold component found in the non-detection stack increases as we move closer to the regions with higher densities of cold gas detections (matching the value of the detection stack used in Section \ref{sec: Stacked Spectrum Detections}), also highlights the unbiased nature of the GASKAP absorption survey in determining regions with higher amounts of cold gas.

The primary component's spin temperature on the other hand remains relatively stable across the region, with the value in each bin staying close to the value of the total stack as shown in Fig. \ref{fig: Spatial Bins Tspin}. Since the primary component traces the peak in emission, which typically accounts for the bulk of cold gas, this result seems to suggest there is little structural variation of the CNM's temperature across this region of the sky. Assuming that the \HI\ absorption lies at the same physical distance as the peak in the weighted mean dust extinction spectra, the average physical distance to each \HI\ absorption location is approximately $240\pm48$\,pc (as shown in Fig. \ref{fig: Edenhofer Average Dust Spectra}). Using this physical distance along with the angular extent of the entire GASKAP Pilot survey of $\sim 25^{\circ}$, the temperature of cold gas within this region of the Milky Way does not vary over a plane-of-sky distance of $\sim100\pm20$\,pc. We note that due to the width of the Local Bubble's shell in the LOS direction, this length scale likely reflects only the plane-of-sky direction and not the depth along the LOS.

The secondary component's lower limit of optical depth exhibits negligible spatial variation from each bin compared to the bulk of the cold gas in the primary components. The optical depth plateaus in the middle bins at a value close to that in the detection stack in Section \ref{sec: Stacked Spectrum Detections} as shown in Fig. \ref{fig: Properties In Each Spatial Bin}. The spin temperature of the secondary component also shows a similar trend as shown in Fig. \ref{fig: Spatial Bins Tspin}. However, due to the lack of sample size, detecting a secondary component in only 2 of the 6 non-detection bins, we cannot make any conclusions on the spatial variation of the secondary component's properties with any confidence.

\section{Summary} 
\label{sec: Summary}

We present analysis using the recent Milky Way GASKAP Phase II Pilot survey's \HI absorption detection sample investigating the effect of multiple components along individual sightlines on resulting stacked spectra and reported measurements such as the spin temperature and optical depth. Taking these effects into consideration, we report the results of stacking the Milky Way \HI absorption and emission spectra from both the detection and non-detections of the GASKAP survey towards the Magellanic system. The main conclusions of this work are as follows:

\begin{itemize}
  \item When stacking multicomponent \HI absorption and emission spectra, any secondary components not centred before stacking create a signal with a peak optical depth lower than the true average value. The FWHM of the component is conversely an upper limit to conserve the total flux of the individual stacked components.
  \item Due to the velocity offset being equal in emission and absorption, the spin temperature from a stack of secondary components is still well correlated with average of the individual components.
  \item Stacking the residual spectra of the 462 GASKAP absorption detection sightlines to improve the total signal-to-noise, we detected gas in \HI absorption with a spin temperature of $\sim1320\pm263$\,K belonging to the UNM.
  \item From stacking the 2240 non-detection GASKAP spectra to find gas hidden within the noise of individual sightlines, we detected the existence of a primary narrow and secondary broad component in \HI\ absorption.
  \item Both the primary and secondary component can be attributed to averaged CNM fields in this region of the sky, with a spin temperature of approximately $98\pm12$\, and $255\pm106$\,K respectively.
  \item When spatially binning these non-detections by their plane-of-sky distance from known regions of cold gas, the primary CNM component varies noticeably in optical depth by almost an order of magnitude, decreasing with distance from the locations of known cold gas.
  \item The change in the primary cold component's optical depth is also reflected in the change in the mean peak extinction density traced by dust over the same lines-of-sight.
  \item The spin temperatures of cold gas components in each bin remain constant and within $1\sigma$ of the spin temperature values from the stack of all 2240 non-detections regardless of their spatial location, over a physical plane-of-sky distance of $\sim100$ pc.
\end{itemize}

In the near-future, the GASKAP survey will observe the same fields discussed in this work with longer exposure times, yielding a higher signal-to-noise ratio and sensitivity. Assuming the same number of sources are found as a conservative estimate, the signal-to-noise of all the absorption stacks throughout this work should improve by almost an order of magnitude. From this, we will then approach a signal-to-noise rivalling that of the stacked 21-SPONGE spectra \citep{Murray+2014,Murray+2018} and will be able to detect higher spin temperature components, possibly even belonging to the WNM.

\section*{Acknowledgements}
This scientific work uses data obtained from Inyarrimanha Ilgari Bundara/the Murchison Radio-astronomy Observatory. We acknowledge the Wajarri Yamaji People as the Traditional Owners and native title holders of the Observatory site. CSIRO’s ASKAP radio telescope is part of the Australia Telescope National Facility (https://ror.org/05qajvd42). Operation of ASKAP is funded by the Australian Government with support from the National Collaborative Research Infrastructure Strategy. ASKAP uses the resources of the Pawsey Supercomputing Research Centre. Establishment of ASKAP, Inyarrimanha Ilgari Bundara, the CSIRO Murchison Radio-astronomy Observatory and the Pawsey Supercomputing Research Centre are initiatives of the Australian Government, with support from the Government of Western Australia and the Science and Industry Endowment Fund.
This research is supported by an Australian Government Research Training Program (RTP) Scholarship.
This research was partially funded by the Australian Government through the Australian Research Council through the award of an Australian Laureate Fellowship (project number FL210100039) to NMc-G.
The authors acknowledge Interstellar Institute’s programme “II6” and the Paris-Saclay University’s Institut Pascal for hosting discussions that nourished the development of the ideas behind this work. 
EDT was supported by the European Research Council (ERC) under grant agreement no. 10104075. 

This research made use of {\fontfamily{qcr}\selectfont Astropy}, a community-developed core
Python package for astronomy \citep{AstropyCollaboration+2013},
{\fontfamily{qcr}\selectfont NumPy} \citep{vanderWalt+2011}, {\fontfamily{qcr}\selectfont matplotlib}, a
Python library for the publication of quality graphics \citep{Hunter+2007},
and {\fontfamily{qcr}\selectfont SciPy} \citep{Virtanen+2020}.

\section*{Data Availability}

The GASKAP-\HI Pilot survey data are available on the CSIRO ASKAP Science Data Archive\footnote{\url{https://research.csiro.au/ casda/}} (CASDA). The non-detection data set and auxiliary data products from this article will be shared on reasonable request to the corresponding author.


\bibliographystyle{mnras}
\bibliography{paper}

\appendix
\section{Toy Model Theoretical Considerations}
\label{sec: Toy Model Theoretical Considerations}

In this section, we use toy models to investigate the biases introduced by mixing complex multiphase emission and absorption spectra through stacking.

\subsection{Single component}
\label{subsec: Single Component}

We first begin our toy model by assuming the simple case of all sightlines containing a single component. Here, every component is perfectly shifted to 0 \kms\ by the peak of the spectrum in emission. However as mentioned in Section \ref{subsec: Velocity Alignment}, this shifting may be inaccurate to $\sim1-2$ \kms\ from the misalignment of absorption and emission peaks due to differing noise levels and sightlines used.

To test how stacking these components affects measurements such as the spin temperature, we set up a toy model with 2000 sightlines, each with an emission and absorption spectrum. In each sightline, a single component was created with a width and amplitudes in brightness temperature and optical depth typical of cold gas in the CNM. The properties for the 2000 sightlines were sampled from Gaussian distributions with the mean value uniformly chosen within $5 \le T_{\mathrm {B,peak}} \le 15$ K, $0.02 \le \tau_{\mathrm {peak}} \le 0.3$, FWHM $\sim5$ \kms\ and standard deviations of $\sigma_{T_{\rm B}} = 4$ K, $\sigma_{\tau_{\rm peak}} = 0.01$ and $\sigma_{\rm FWHM} = 0.2$. The central velocities of these components is centred around 0 \kms, with a slight deviation to account for the $1-2$ \kms\ deviation between emission and absorption components found in reality (a central velocity dispersion of $\sim0.5$ in the toy model). Stacking these spectra, we then calculate the spin temperature, $T_{s}$ of the stacked component via:
\begin{align}
    T_{s} = \frac{T_{\rm B,\ peak}}{1-e^{-\tau_{\rm peak}}}
\end{align}

Alongside this, we also compute the spin temperature of the individual sightlines and take the average to create a spin temperature that should be measured. Repeating this scenario 200 times (using a different sampled mean value of $T_{\rm B, peak}$ and $\tau_{\rm peak}$ for each iteration), we find the spin temperature of the stack is well correlated with the average spin temperature of the individual sightlines (figure not shown).

\subsection{Multiple components}
\label{subsec: Multiple Components}

As discussed in Section \ref{subsec: Velocity Alignment}, the most sensitive detection sources contain multiple components: a narrow primary component aligned with the peak in the corresponding emission profile, and a broader secondary component offset in velocity.

The primary component tied to the peak in emission will be aligned well at 0 \kms\ after shifting, whereas the additional peaks will merge into a broader secondary component.

To explore these effects of stacking multicomponent spectra, we create a similar toy model to that of Section \ref{subsec: Single Component}. We create 2000 mock sightlines, each with an absorption and emission spectrum. We create primary components similar to the previous section. In addition, we add a broader component. This secondary component has a width and brightness temperature and optical depth amplitude randomly sampled within typical values of warmer CNM and colder UNM gas. The properties for the 2000 sightlines were sampled from Gaussian distributions with the mean value uniformly chosen within $3 \le T_{\mathrm {B,peak}} \le 15$ K, $0.02 \le \tau_{\mathrm {peak}} \le 0.03$, FWHM $\sim15$ \kms\ and standard deviations of $\sigma_{T_{\rm B}} = 4$ K, $\sigma_{\tau_{\rm peak}} = 0.01$ and $\sigma_{\rm FWHM} = 0.6$ respectively. The central velocity of this component is randomly sampled over a normal distribution centred on 0 \kms\ but with a dispersion of $\sim15$ \kms, simulating the potential of a non-zero central velocity of these secondary components after shifting.

\begin{figure}
    \includegraphics[width=\columnwidth]{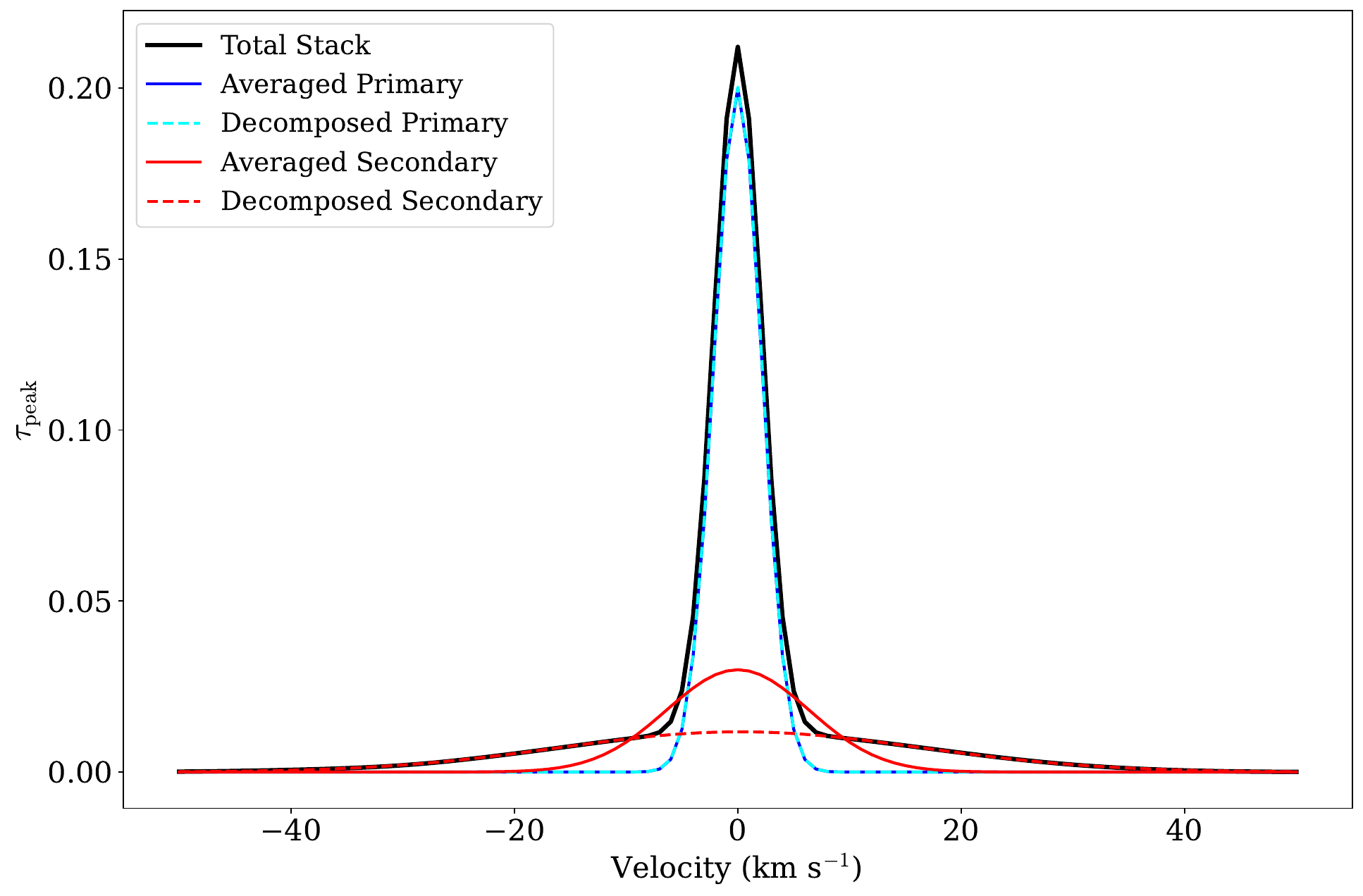}
    \caption{Comparison of the primary and secondary component from Gaussian decomposition of a toy model stacked absorption spectra with the separated stacks of the individual primary and secondary components. The total stack (black) is produced as described in Section \ref{subsec: Multiple Components}. We note the primary component in the stack (cyan-dashed) is identical to the separately averaged primary component (solid blue). The secondary component (red-dashed) is shallower and has a greater FWHM than the separately averaged secondary component (solid red) where each component was centered to 0 \kms.}
    \label{fig: Broadening and Peak Test}
\end{figure}

Fig. \ref{fig: Broadening and Peak Test} demonstrates the effects of stacking these multicomponent absorption sightlines featuring a primary component, shifted correctly to 0\,\kms, and a broader secondary component whose central velocity is chosen from a normal distribution with a deviation of 15 \kms. The values for the peaks and widths of the two components were chosen based on the parameter space mentioned earlier. The primary components, being correctly centred, produce an averaged component after Gaussian decomposition equivalent to an average of the individual primary components. However, due to the velocity variation of the secondary components, the average secondary component found after Gaussian decomposition varies drastically in amplitude and width from the average of the individual sightlines of the secondary components if they were all centred at 0 \kms.

\begin{figure}
    \includegraphics[width=\columnwidth]{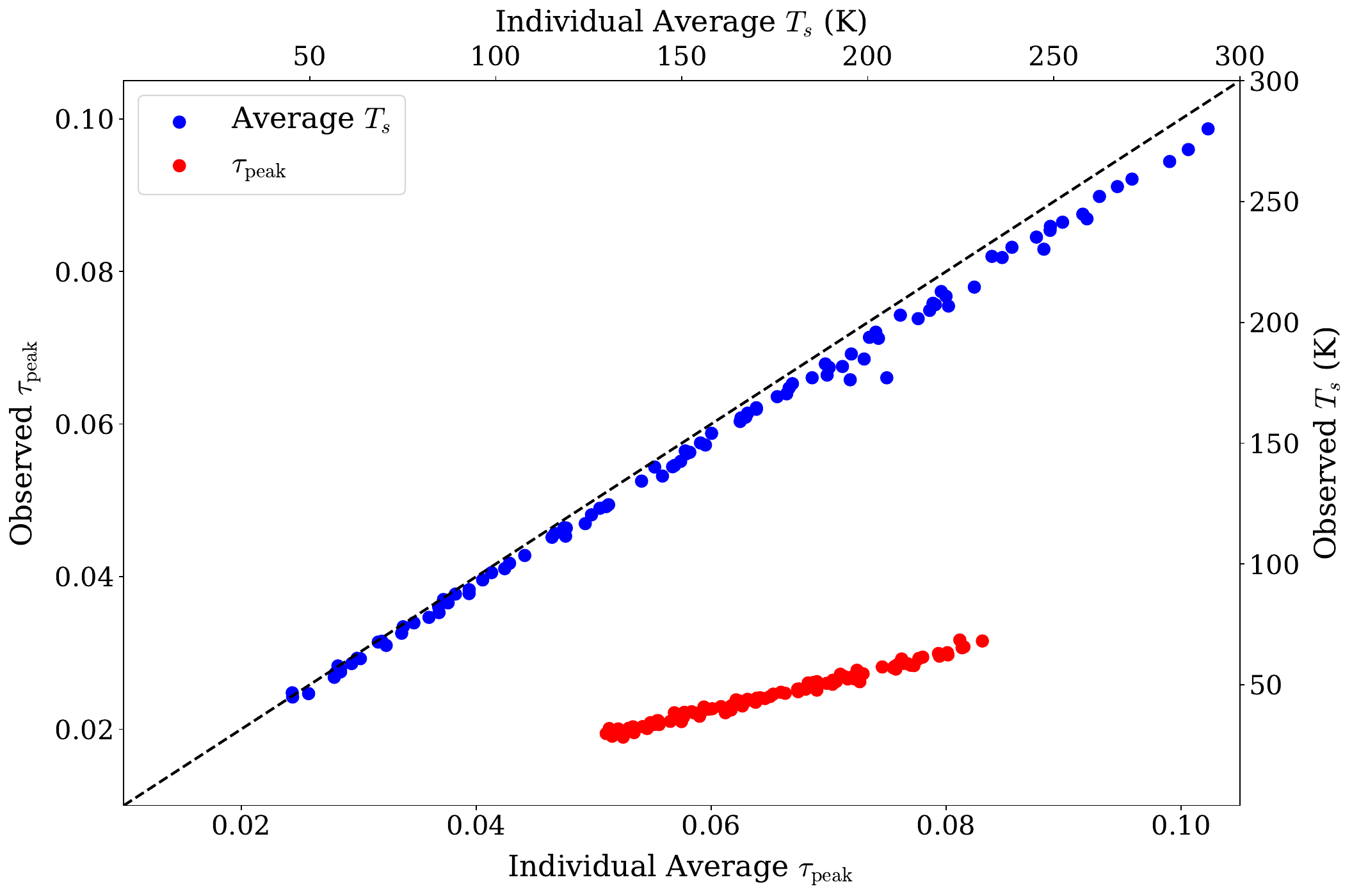}
    \caption{Comparison of observed peak optical depth (red) and average spin temperature (blue) of the secondary components from 100 runs of stacking 2000 two-component spectra, as described in Section \ref{subsec: Multiple Components}. The peak brightness temperature is not displayed but shows a similar trend to the peak optical depth due to the velocity offsets being the same in both emission and absorption. The observed peak optical depth is always less than the average peak optical depth of the individual sightlines' secondary components. However, the observed spin temperature of the secondary component correlates well with the average spin temperature of each sightline's secondary component.}
    \label{fig: Toy Model Properties Observed vs True}
\end{figure}

Repeating this experiment 100 times (with a different sampled mean value of $T_{\rm B, peak}$ and $\tau_{\rm peak}$ for each iteration within the ranges mentioned above), we compared the measured amplitudes of the components with that of the average amplitude of the components in emission and absorption in individual sightlines. This comparison is shown in Fig. \ref{fig: Toy Model Properties Observed vs True}. Only the secondary component's properties are displayed since the primary component's values agree with the average of individual primary components due to the correct centring of the features prior to stacking. For the secondary components however, the measured amplitudes are always lower. This is because the total flux is conserved in the average regardless of the Gaussian's central velocities. Increasing the velocity spread of the Gaussians broadens the final stack, and to compensate the averaged amplitude must diminish. Therefore, regardless of the severity of the range of central velocities of secondary components in individual sightlines, the peak of the secondary component will always be lower and its width always greater than the average of the individual sightline values.

Despite this lower limit, the stacked secondary components' amplitudes are still linearly correlated with the average values. Since the central velocities of the secondary components are approximately equal in emission and absorption, the overall effect of the velocity dispersion on the stack is equivalent in emission and absorption. As a result, the gradient of the linear correlation of measured peak brightness temperature and optical depth with the individual sightline average is the same.

When calculating the spin temperature of the secondary component through this toy model, we find that these gradients cancel out through the division of emission and absorption amplitudes. The spin temperature of the secondary component is therefore well correlated with the average of the individual secondary components' spin temperatures, as shown in Fig. \ref{fig: Toy Model Properties Observed vs True}.

%


\bsp	
\label{lastpage}
\end{document}